\documentclass[12pt]{elsart}
\usepackage{times}
\usepackage{epsfig}
\usepackage{graphics}
\begin{document}

\newcommand{\re}{\mathop{\mathrm{Re}}}
\newcommand{\im}{\mathop{\mathrm{Im}}}
\newcommand{\I}{\mathop{\mathrm{i}}}
\newcommand{\D}{\mathop{\mathrm{d}}}
\newcommand{\E}{\mathrm{e}}

\def\lambar{\lambda \hspace*{-5pt}{\rule [5pt]{4pt}{0.3pt}} \hspace*{1pt}}

\begin{Large}
\textbf{DEUTSCHES ELEKTRONEN-SYNCHROTRON}

\textbf{\large{in der HELMHOLTZ-GEMEINSCHAFT}\\}
\end{Large}

{\large DESY 06-137

August 2006}

\bigskip
\bigskip
\bigskip

\begin{frontmatter}

\journal{Optics Communications}

\date{}

\title{
Coherence properties of the radiation from X-ray free electron laser
}

\author{E.L.~Saldin},
\author{E.A.~Schneidmiller},
\author{M.V.~Yurkov}

\address{Deutsches Elektronen-Synchrotron (DESY),
Notkestrasse 85, D-22607 Hamburg, Germany}

\begin{abstract}

We present a comprehensive analysis of coherence properties of the radiation from X-ray free electron laser (XFEL). We consider practically important case when XFEL is optimized for maximum gain. Such an optimization allows to reduce significantly parameter space. Application of similarity techniques to the results of numerical simulations allows to present all output characteristics of the optimized XFEL as functions of the only parameter, ratio of the emittance to the radiation wavelength, $ \hat{\epsilon } = 2\pi \epsilon/\lambda $. Our studies show that optimum performance of the XFEL in terms of transverse coherence is achieved at the value of the parameter $\hat{\epsilon }$ of about unity. At smaller values of $\hat{\epsilon }$ the degree of transverse coherence is reduced due to strong influence of poor longitudinal coherence on a transverse one. At large values of the emittance the degree of transverse coherence degrades due to poor mode selection. Comparative analysis of existing XFEL projects, European XFEL, LCLS, and SCSS is presented as well.

\end{abstract}




\end{frontmatter}

\baselineskip 18pt

\clearpage

\thispagestyle{empty}

$\mbox{}$

\clearpage

\setcounter{page}{1}

\section{Introduction}

Free electron lasing at wavelengths shorter than ultraviolet  can
be achieved with a single-pass, high-gain FEL amplifier. Because a lack of
powerful, coherent seeding sources short-wavelength FEL amplifiers work in so
called Self-Amplified Spontaneous Emission (SASE) mode when amplification
process starts from shot noise in the electron beam
\cite{kondratenko-1980,derbenev-xray-1982,pellegrini}. Present level of
accelerator and FEL techniques holds potential for SASE FELs to generate
wavelengths as short as 0.1~nm \cite{euro-xfel-tdr,lcls-cdr,scss-cdr}.

Experimental realization of X-ray FELs (XFELs) developed very rapidly during
last decade. The first demonstration of the SASE FEL mechanism took place in
1997 in the infrared wavelength range \cite{pellegrini-sase}. In September of
2000, a group at Argonne National Laboratory (ANL) became the first to
demonstrate saturation in a visible (390 nm) SASE FEL \cite{leutl-sat}. In
September 2001, a group at DESY (Hamburg, Germany) has demonstrated lasing to
saturation at 98 nm \cite{ttf-sat-prl,ttf-sat-epj}. In June 2006 saturation has
been achieved at 13 nm, the shortest wavelength ever generated by FELs. The
experimental results have been achieved at FLASH ("F"ree-Electron-"LAS"er in
"H"amburg). Regular user operation of FLASH started
in 2005 \cite{vuvfel-exp}. Currently FLASH produces GW-level, laser-like VUV
radiation pulses with 10 to 50~fs duration in the wavelength range 13-45 nm.
After the energy upgrade of the FLASH linac to 1~GeV planned in 2007, it will
be possible to generate wavelengths down to 6~nm.

Recently the German government, encouraged by these results, approved funding a
hard X-ray SASE FEL user facility -- the European X-Ray Free Electron Laser
\cite{euro-xfel-tdr}. The US Department of Energy (DOE) has given SLAC the
goahead for the engineering design of the  Linac Coherent Light Source (LCLS)
to be constructed at SLAC \cite{lcls-cdr}.  These devices should produce
100 fs X-ray pulses with over 10 to 100 GW of peak power. The main difference
between projects is the linear accelerator, an existing room temperature linac
for LCLS at SLAC, and future superconducting linac for the European XFEL. The
XFEL based on superconducting accelerator technology will make possible not
only a jump in a peak brilliance by ten orders of magnitude, but also increase
by five orders of magnitude in average brilliance. The LCLS and European XFEL
projects are scheduled to start operation in 2009 and 2013, respectively.

In the X-ray FEL the radiation is produced by the electron beam during
single-pass of the undulator
\cite{kondratenko-1980,derbenev-xray-1982,pellegrini}. The amplification
process starts from the shot noise in the electron beam. Any random
fluctuations in the beam current correspond to a modulation of the
beam current at all frequencies simultaneously. When the electron beam enters
the undulator, the presence of the beam modulation at frequencies close to the
resonance frequency initiates the process of radiation. The FEL collective
instability in the electron beam produces an exponential growth (along the
undulator) of the modulation of the electron density on the scale of undulator
radiation wavelength. The fluctuations of current density in the electron beam
are uncorrelated not only in time but in space, too. Thus, a large number of
transverse radiation modes are excited when the electron beam enters the
undulator. These radiation modes have different gain. As undulator length
progresses, the high gain modes start to predominate more and more. For enough
long undulator, the emission will emerge in a high degree of transverse
coherence. An intensity gain in excess of $10^6-10^7$ is obtained in the
saturation regime. At this level, the shot noise of the electron beam is
amplified up to complete micro-bunching, and all electrons radiate almost in
phase producing powerful, coherent radiation.

Understanding of coherence properties of the radiation from SASE FEL is of
great practical importance. Properties of the longitudinal coherence
have been studied in
\cite{1d1,1d2,1d3,bon-prl-1994,pierini-bill,1d6,statistics-oc}. It has been
found that the coherence time increases first, reaches
maximum value in the end of the linear high gain regime and then drops when
amplification process enters nonlinear stage \cite{statistics-oc}. The first
analysis of the problem of transverse coherence has been performed in
\cite{trcoh-oc}. The problem of start-up from the shot noise has been studied
analytically and numerically for the linear stage of amplification. It has been
found that the process of formation of transverse coherence is more complicated
than that given by naive physical picture of transverse mode selection. Namely,
even after finishing the transverse mode selection process the degree of
transverse coherence of the radiation from SASE FEL visibly differs from the
unity.  This is consequence of the interdependence of the longitudinal and
transverse coherence. The SASE FEL has poor longitudinal coherence which
develops slowly with the undulator length thus preventing a full transverse
coherence. First studies of the evolution of transverse coherence in the
nonlinear regime of SASE FEL operation have been performed in
\cite{trcoh-nima}. It has been found that similarly to the coherence time, the
degree of transverse coherence reaches maximum value in the end of the linear
regime. Further increase of the undulator length leads to its decrease. Despite
output power of the SASE FEL grows continuously in the nonlinear regime,
maximum brilliance of the radiation is achieved in the very beginning of the
nonlinear regime. Due to a lack of computing power available at that time we
limited our study with a specific numerical example just illustrating the
general features of coherence properties of the radiation produced by the SASE
FEL operating in the nonlinear regime.

In this paper we present general analysis of the coherence properties
(longitudinal and transverse) of the radiation from SASE FEL. The results have
been obtained with time-dependent, three-dimensional FEL simulation code FAST
\cite{fast} performing simulation of the FEL process with actual number of
electrons in the beam. Using similarity techniques we present universal
dependencies for the main characteristics of the SASE FEL covering all
practical range of X-ray FELs.

\section{Basic relations}

Design of the focusing system of XFEL assumes nearly uniform focusing of the
electron beam in the undulator, so we consider axisymmetric model of the
electron beam. It is assumed that transverse distribution function of the
electron beam is Gaussian, so rms transverse size of matched beam is $\sigma =
\sqrt{\epsilon \beta }$ ,where $\epsilon = \epsilon_{\mathrm{n}}/\gamma$ is rms
beam emittance and $\beta $ is focusing beta-function. An important feature of
the parameter space of XFEL is that the space charge field does not influence
significantly on the FEL process and calculation of the FEL process can be
performed by taking into account diffraction effects, the energy spread in the
electron beam, and effects of betatron motion only. In the framework of the
three-dimensional theory operation of the FEL amplifier is described by the
following parameters:
the diffraction parameter $B$,
the energy spread parameter $\hat{\Lambda }^{2}_{\mathrm{T}}$,
and the betatron motion parameter $\hat{k}_{\beta }$
\cite{book,eigen-general}:

\begin{eqnarray}
\nonumber \\
B & =& 2 \Gamma \sigma^2 \omega/c \ ,
\nonumber \\
\hat{k}_{\beta} & = & 1/(\beta \Gamma ) \ ,
\nonumber \\
\hat{\Lambda }^{2}_{\mathrm{T}} & = &
(\sigma _{\mathrm{E}}/{\mathcal E})^2/\rho^2  \ ,
\label{eq:reduced-parameters}
\end{eqnarray}

\noindent where
$\Gamma = \left[ I \omega^2 \theta_{\mathrm{s}}^2 A_{\mathrm{JJ}}^2/
(I_{\mathrm{A}}
c^2 \gamma_{\mathrm{z}}^2 \gamma ) \right]^{1/2}$ is the gain parameter
and $\rho = c\gamma ^{2}_{\mathrm{z}}\Gamma /\omega $ is the efficiency
parameter\footnote{Note that it differs from 1-D definition by the factor
$B^{1/3}$ \cite{book}.}. When describing shot noise in the electron beam, one
more parameter appears, the number of electrons on the coherence length,
$N_{\mathrm{c}} = I/(e\omega \rho )$.
The following notations are used here: $I$ is the beam current,
$\omega = 2\pi c/\lambda$ is the frequency of the electromagnetic wave,
$\theta_{\mathrm{s}}=K_{\mathrm{rms}}/\gamma$, $K_{\mathrm{rms}}$ is the rms
undulator parameter, $\gamma$ is relativistic factor, $\gamma ^{-2}_{z} =
\gamma ^{-2}+ \theta ^{2}_{\mathrm{s}}$, $k_{\mathrm{w}} = 2\pi /\lambda
_{\mathrm{w}}$ is the undulator wavenumber, $I_{\mathrm{A}} =$ 17 kA is the
Alfven current, $A_{\mathrm{JJ}} = 1$ for helical undulator and
$A_{\mathrm{JJ}} = J_0(K_{\mathrm{rms}}^2/2(1+K_{\mathrm{rms}}^2)) -
J_1(K_{\mathrm{rms}}^2/2(1+K_{\mathrm{rms}}^2))$ for planar undulator.  Here
$J_0$ and $J_1$ are the Bessel functions of the first kind. The energy spread
is assumed to be Gaussian with rms deviation $\sigma _{\mathrm{E}}$.

The amplification process in the FEL amplifier passes two stages, linear and
nonlinear. The linear stage lasts over significant fraction of the undulator
length (about 80\%), and the main target for XFEL optimization is the field
gain length. In the linear high-gain limit the radiation emitted by the
electron beam in the undulator can be represented as a set of modes:

\begin{eqnarray}
E_{\mathrm{x}} + iE_{\mathrm{y}} & = & \int \mathrm{d}\omega
\exp [i\omega (z/c-t)]
 \times \sum \limits _{n,m}
A_{nm}(\omega , z) \Phi_{nm}(r,\omega )
\exp [ \Lambda _{nm}(\omega )z + in\phi ] \ .
\label{eq:modes}
\end{eqnarray}

\noindent When amplification takes place, the mode configuration in the
transverse plane remains unchanged while the amplitude grows exponentially with
the undulator length. Each mode is characterized by the eigenvalue $\Lambda
_{nm}(\omega )$ and the field distribution eigenfunction $\Phi _{nm}(r,\omega
)$ in terms of transverse coordinates. At sufficient undulator length
fundamental TEM$_{00}$ mode begins to give main contribution to the total
radiation power. Thus, relevant value of interest for XFEL optimization is the
field gain length of the fundamental mode, $L_{\mathrm{g}} = 1/\re ( \Lambda
_{00})$, which gives good estimate for expected length of the undulator needed
to reach saturation, $L_{\mathrm{sat}} \simeq 10\times L_{\mathrm{g}}$.
Optimization of the field gain length is performed by means of numerical
solution of the corresponding eigenvalue equations taking into account all the
effects (diffraction, energy spread and emittance) \cite{eigen-general,ming}.
Computational possibilities of modern computers allows to trace complete
parameter space of XFEL (which in fact is 11-dimensional). From practical point
of view it is important to find an absolute minimum of the gain length
corresponding to optimum focusing beta function. For this practically important
case the solution of the eigenvalue equation for the field gain length of the
fundamental mode and optimum beta function are rather accurately approximated
by \cite{xfel-fit}:

\begin{eqnarray}
L_{\mathrm{g}} & = &
1.67 \left(\frac{I_A}{I} \right)^{1/2}
\frac{(\epsilon_n \lambda_{\mathrm{w}})^{5/6}}
{\lambda ^{2/3}} \ \frac{(1+K^2)^{1/3}}{K A_{JJ}} ( 1 + \delta ) \ ,
\nonumber \\
\beta_{\mathrm{opt}} & \simeq & 11.2 \left(\frac{I_A}{I} \right)^{1/2}
\frac{\epsilon_n^{3/2}
\lambda_{\mathrm{w}}^{1/2}}
{\lambda K A_{JJ}} \ (1+8\delta)^{-1/3} \ ,
\nonumber \\
\delta & = &
131 \ \frac{I_A}{I} \ \frac{\epsilon_n^{5/4}}
{\lambda ^{1/8} \lambda_{\mathrm{w}}^{9/8}} \
\frac{\sigma_{\gamma}^2}{(K A_{JJ})^2 (1+K^2)^{1/8}}
\ ,
\label{eq:lg}
\end{eqnarray}

\noindent where $\sigma_{\gamma}=\sigma_{\mathrm{E}}/m_{\mathrm{e}}c^2$.
Accuracy of this fit is better than 5\% in the range of
parameter $\hat{\epsilon } = 2 \pi \epsilon/\lambda$ from 1 to 5.

Equation (\ref{eq:lg}) demonstrates clear interdependence of physical
parameters defining operation of the XFEL. Let us consider the case of
negligibly small energy spread. Under this condition
diffraction parameter $B$ and parameter of betatron oscillations,
$\hat{k}_{\beta}$ are functions of the only parameter
$\hat{\epsilon }$:

\begin{equation}
B \simeq 12.5 \times \hat{\epsilon }^{5/2} \ , \qquad
\hat{k}_{\beta} = 1/(\beta \Gamma )
\simeq 0.158 \times \hat{\epsilon }^{3/2} \ .
\label{eq:bopt}
\end{equation}

\noindent FEL equations written down in the dimensionless form involve an
additional parameter $N_{\mathrm{c}}$ defining the initial conditions for the
start-up from the shot noise. Note that the dependence of output
characteristics of the SASE FEL operating in saturation is slow, in fact
logarithmic in terms of $N_{\mathrm{c}}$. Thus, we can conclude that with
logarithmic accuracy in terms of $N_{\mathrm{c}}$, characteristics of the SASE
FEL written down in a normalized form are functions of the only parameter
$\hat{\epsilon}$.

\section{An approach for numerical simulations}

Rigorous studies of the nonlinear stage of amplification is possible only with
numerical simulation code. Typically FEL codes use an artificial ensemble of
macroparticles for simulating the FEL process when one macroparticle
represents large number of real electrons. Thus, a natural question arises if
macroparticle phase space distributions are identical to those of actual
electron beam at all stages of amplification. Let us trace typical procedure
for preparation of an artificial ensemble \cite{fawley-loading,genesis}. The
first step of particle loading consists in a quasi-uniform distribution of the
macroparticles in the phase space. At this stage an ensemble of particles with
random distribution is generated which occupies a fraction of the phase space.
Then this ensemble is copied on the other parts of the phase space to provide
pseudo-uniform loading of the phase space. Pseudo-uniformity means that initial
microbunching at the fundamental harmonic (or for several harmonics) is equal
to zero. Also, phase positions of the mirrored particles are correlated such
that microbunching does not appear due to betatron oscillations, or due to the
energy spread. Finally, artificial displacements of the macroparticles are
applied to provide desired (in our case gaussian) statistics of microbunching
at the undulator entrance. We note that it is not evident that such an
artificial ensemble reflects actual physical situation for a short wavelength
SASE FEL. Let us consider an example of the SASE FEL operating at the radiation
wavelength of 0.1~nm. With the peak current of 5~kA we find that the number of
electrons per wavelength is about $10^{4}$. On the other hand, it is well known
that properties of an artificial ensemble (even at the first step of
pseudo-uniform loading) converge very slowly to the model of continuous media.
In fact, even with the number of macroparticles per radiation wavelength
$6.4\times 10^{4}$ the FEL gain still visibly deviates from the target value.
Introducing of an artificial noise makes situation with the quality of an
ensemble preparation even more problematic. The only way to test the quality of
an artificial ensemble is to perform numerical simulations with actual number
of electrons in the beam. We constructed such a version of three-dimensional,
time-dependent FEL simulation code FAST \cite{fast}. Comparison of the results
with direct simulations of the electron beam and with an artificial
distributions has shown that artificial ensembles are not adequate to the
problem. Artificial effects are pronouncing especially when calculating such
fine features as transverse correlation functions. Thus, all the simulations
presented in this paper have been performed with code FAST using actual number
of electrons in the beam.

\section{General overview of the properties of the radiation form SASE FEL}

\begin{figure}[b]

\hspace*{0.05\textwidth}
\includegraphics[width=0.23\textwidth]{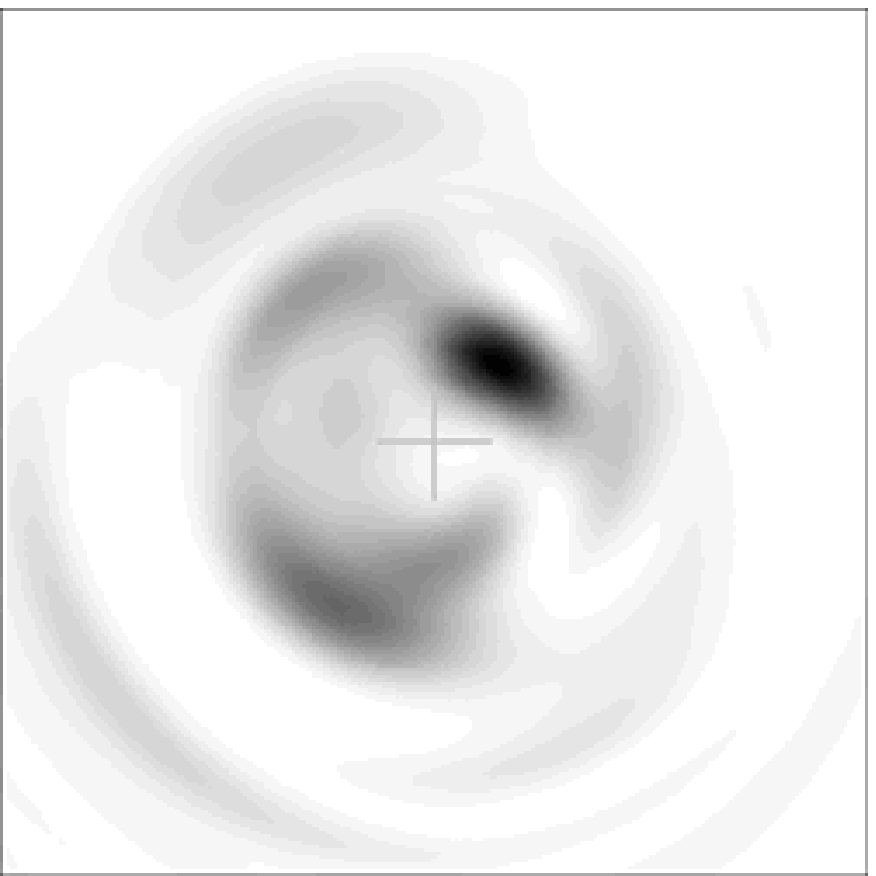}
\hspace*{0.1\textwidth}
\includegraphics[width=0.23\textwidth]{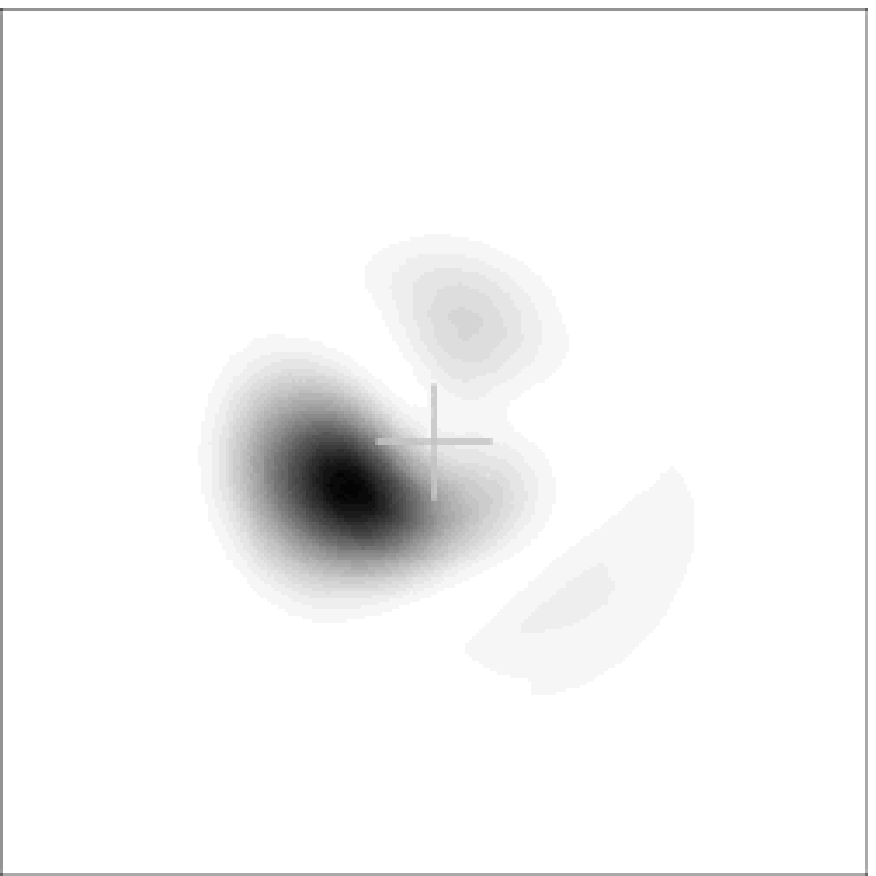}
\hspace*{0.1\textwidth}
\includegraphics[width=0.23\textwidth]{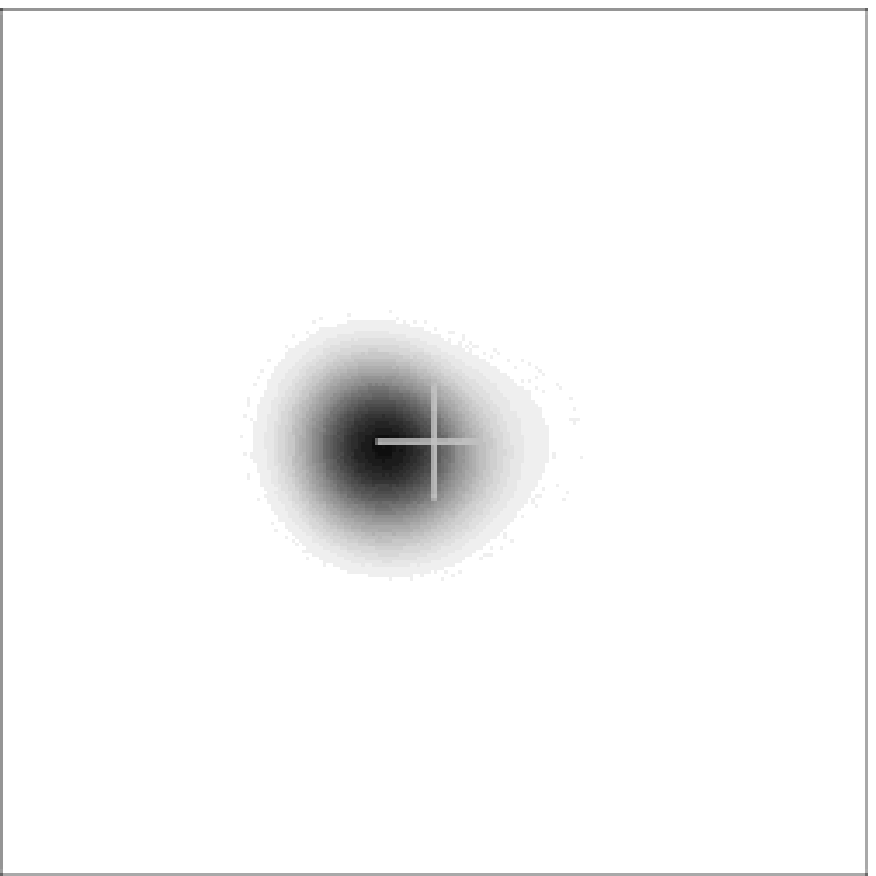}

\vspace*{5mm}

\includegraphics[width=0.32\textwidth]{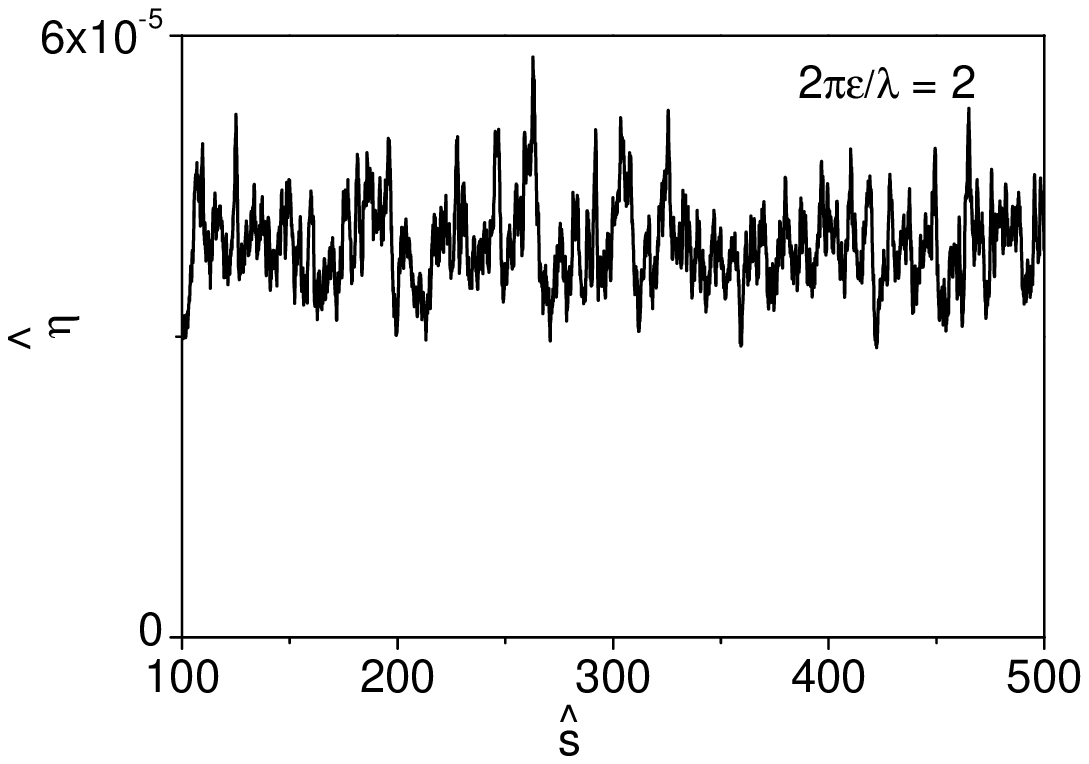}
\includegraphics[width=0.32\textwidth]{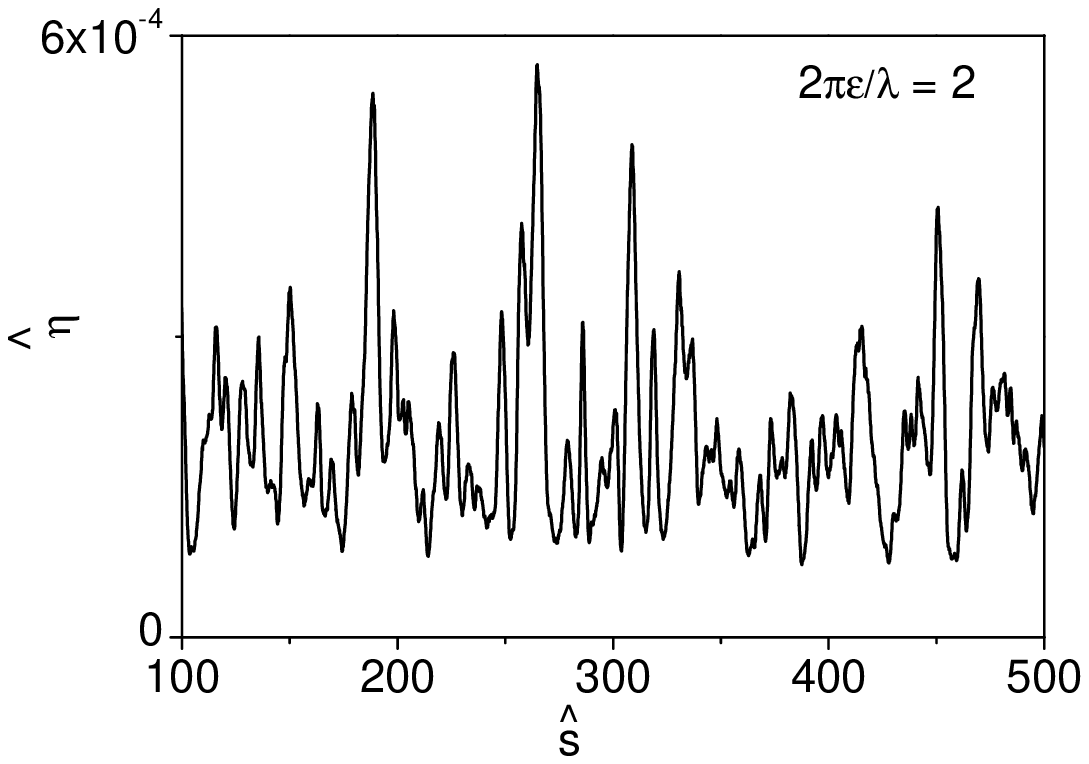}
\includegraphics[width=0.32\textwidth]{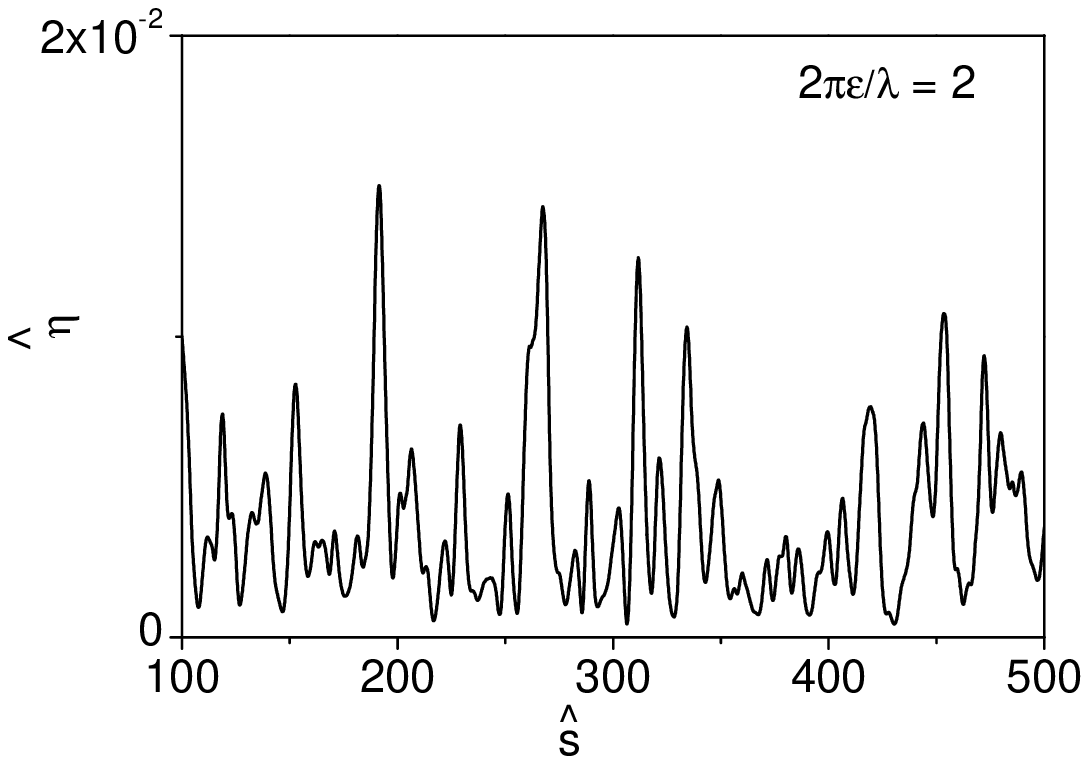}

\caption{Evolution of the power density distribution in a slice
(top) and normalized power in a radiation
pulse (bottom) versus $\hat{s}=\rho\omega_0 (z/\bar{v}_z-
t)$ for the reduced lengths $\hat{z} =
18.4$, 31.6, and 44.8 (left,
middle, and right plots, respectively).
Here $\hat{\epsilon} = 2$.
Crosses show geometrical center of the radiation beam.
}

\label{fig:slice-along-z-em2}
\end{figure}

The result of each simulation run contains an array of complex amplitudes
$\tilde{E}$ for electromagnetic fields on a three-dimensional mesh. At the next
stage of the numerical experiment the data arrays are handled with
postprocessor codes to calculate different characteristics of the radiation.
However, as the first step it is worthwhile to obtain qualitative analysis
of the object under study. The plots in an upper row of
Fig.~\ref{fig:slice-along-z-em2} show evolution of the power density
distribution, $I = |\tilde{E}|^2$, in a slice of the radiation pulse along the
undulator. We see that due to the start-up of amplification process from the
shot noise many transverse radiation modes are excited when electron beam
enters the undulator. Mode selection process (\ref{eq:modes}) serves as a
filter for selection of the fundamental radiation mode having maximum gain.

\begin{figure}[tb]

\includegraphics[width=0.49\textwidth]{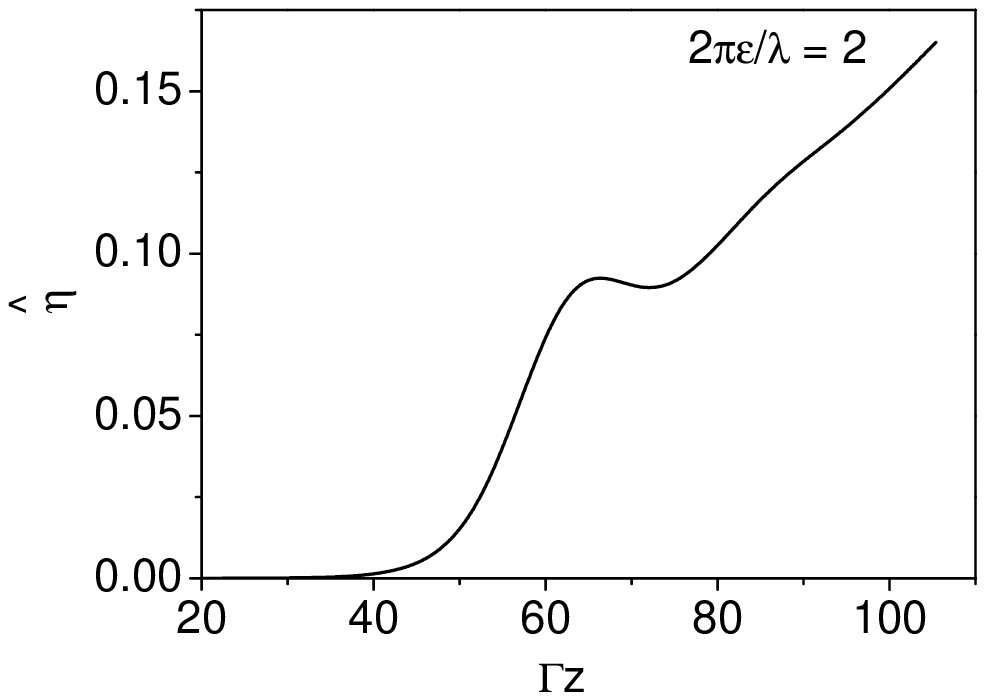}
\includegraphics[width=0.49\textwidth]{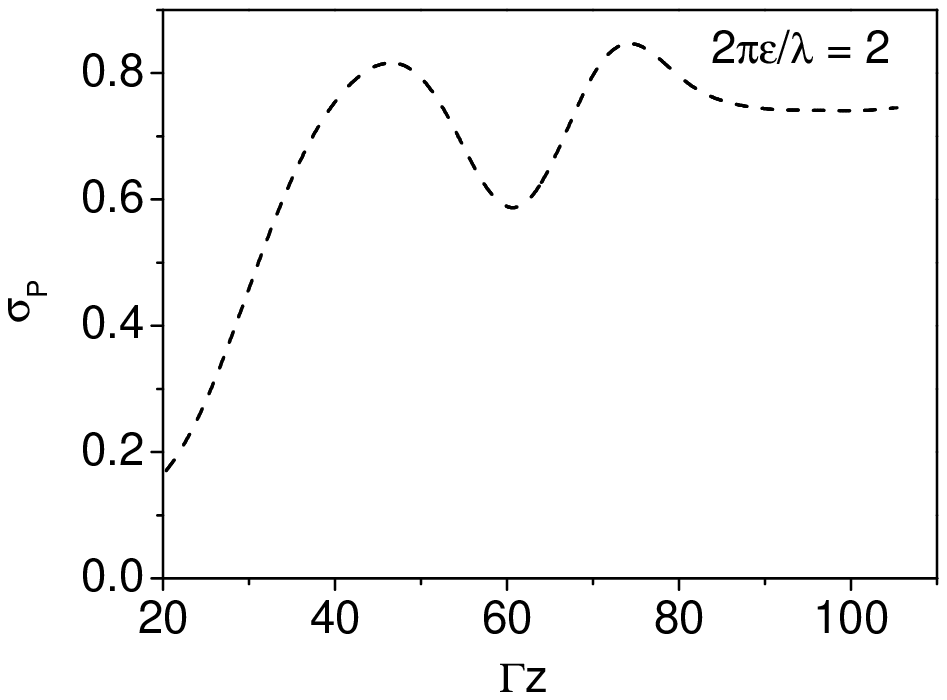}

\caption{Averaged normalized efficiency, $\hat{\eta }$, and
normalized rms deviation of instantaneous radiation power,
$\sigma _{\mathrm{P}}$
along the normalized undulator length, $\hat{z} = \Gamma z$.
Here $\hat{\epsilon } = 2$.
}
\label{fig:pzem2}


\includegraphics[width=0.49\textwidth]{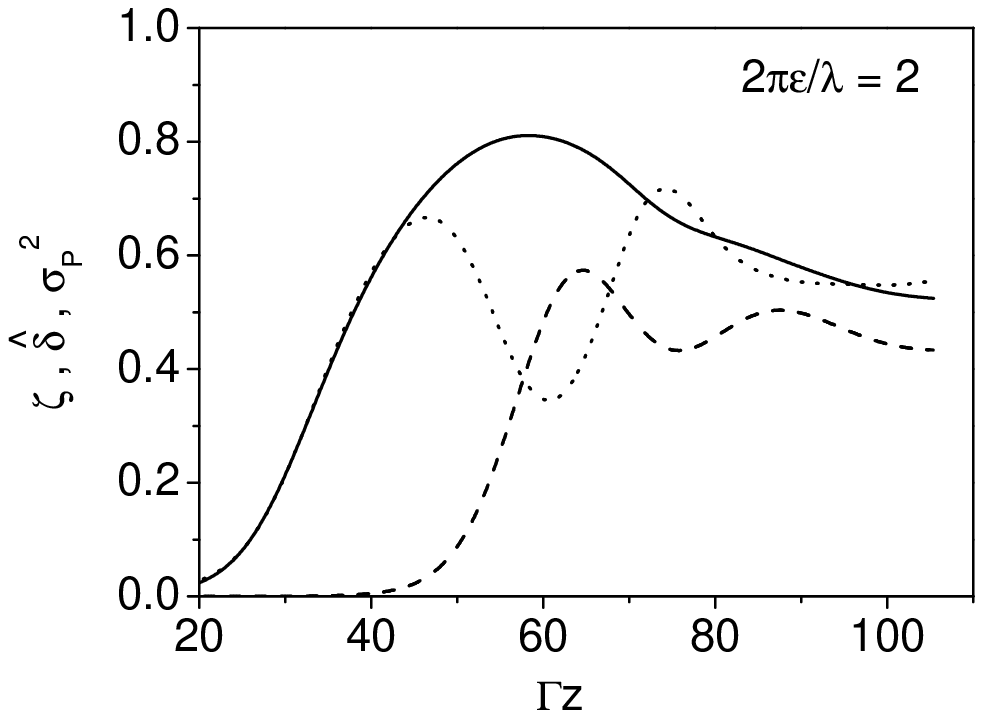}
\includegraphics[width=0.49\textwidth]{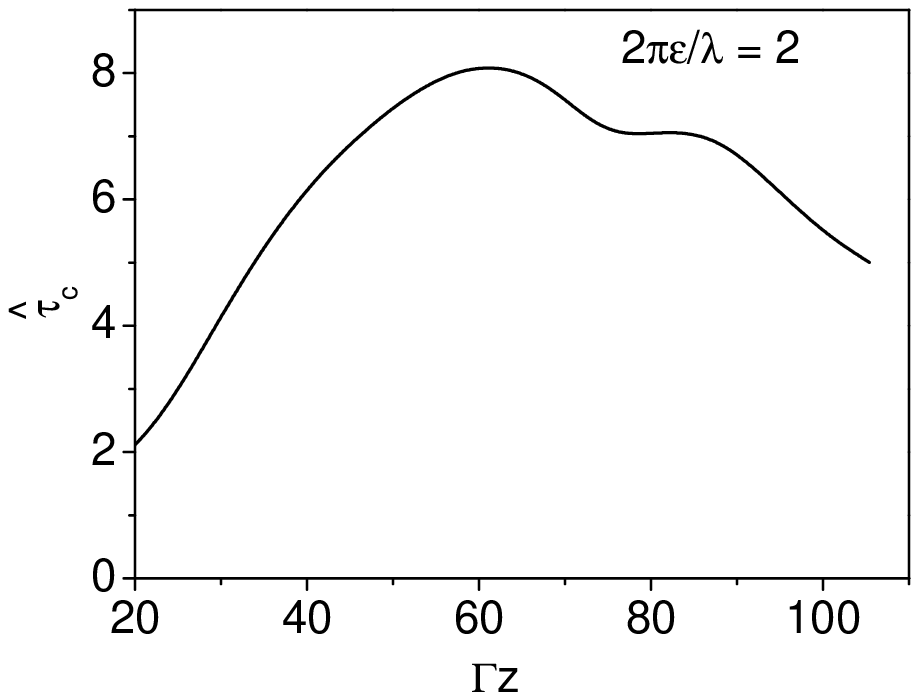}

\caption{
Right plot: normalized coherence time $\hat{\tau }_{\mathrm{c}}$,
along the normalized undulator length, $\hat{z} = \Gamma z$.
Left plot:
degree of transverse coherence, $\zeta $ (solid line), and normalized
degeneracy parameter
of the radiation, $\hat{\delta }$ (dashed line),
along the normalized undulator length, $\Gamma z$.
Dotted line shows $\sigma _{\mathrm{P}}^2$ (see Fig.~\ref{fig:pzem2}).
Here $\hat{\epsilon } = 2$.
}
\label{fig:tcem2}
\end{figure}

Integration of the power density over transverse cross section of the
photon beam gives us instantaneous radiation power,
$P \propto \int I \D \vec{r}_{\perp}$.
Evolution of temporal structure of the radiation power along the
undulator is traced in Fig.~\ref{fig:slice-along-z-em2} in terms of normalized
radiation power $\hat{\eta} = P /(\rho P_{\mathrm{b}})$ where $P_{\mathrm{b}} =
\gamma mc^2I/e$ is the electron beam power. Averaging of the radiation power
along the pulse gives us averaged radiation power. Evolution of normalized
averaged power $\langle \hat{\eta} \rangle $ along  normalized undulator length
$\hat{z} = \Gamma z$ is shown in Fig.~\ref{fig:pzem2}. Note that the radiation
produced by SASE FEL operating in the linear regime holds properties of
completely chaotic polarized light \cite{statistics-oc} -- a statistical object
well described in the framework of statistical optics \cite{goodman}. This is
simple consequence of the fact that the shot noise in the electron beam is a
Gaussian random process. The FEL amplifier, operating in the linear regime, can
be considered as a linear filter which does not change the statistics of the
signal. As a result, we can define general statistical properties of the output
radiation without any calculations.  For instance, in the case of the SASE FEL
the real and imaginary parts of the slowly varying complex amplitudes of the
electric field of the electromagnetic wave, $\tilde{E}$ , have a Gaussian
distribution, the instantaneous power density, $I = |\tilde{E}|^2$,
fluctuates in accordance with the negative exponential distribution (see
Fig.~\ref{fig:hprobem2}):

\begin{equation}
p(I) = \frac{1}
{\langle I \rangle }
\exp\left(-\frac{I}
{\langle I \rangle }\right) \ .
\label{neg-exp-1}
\end{equation}

\noindent Due to the start-up of amplification process from the shot noise many
transverse radiation modes are excited when electron beam enters the undulator.
For gaussian random process any integral of the power density, for
example, radiation power $P$, fluctuates in accordance with the gamma
distribution:

\begin{figure}[tb]
\includegraphics[width=0.32\textwidth]{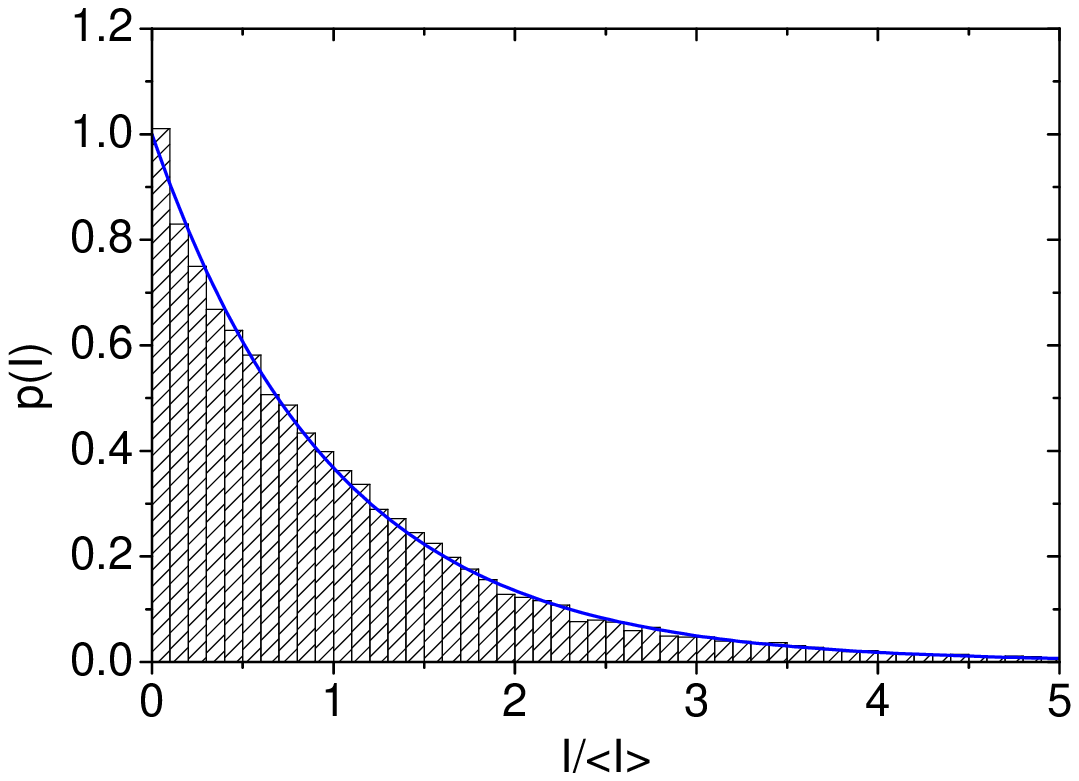}
\includegraphics[width=0.32\textwidth]{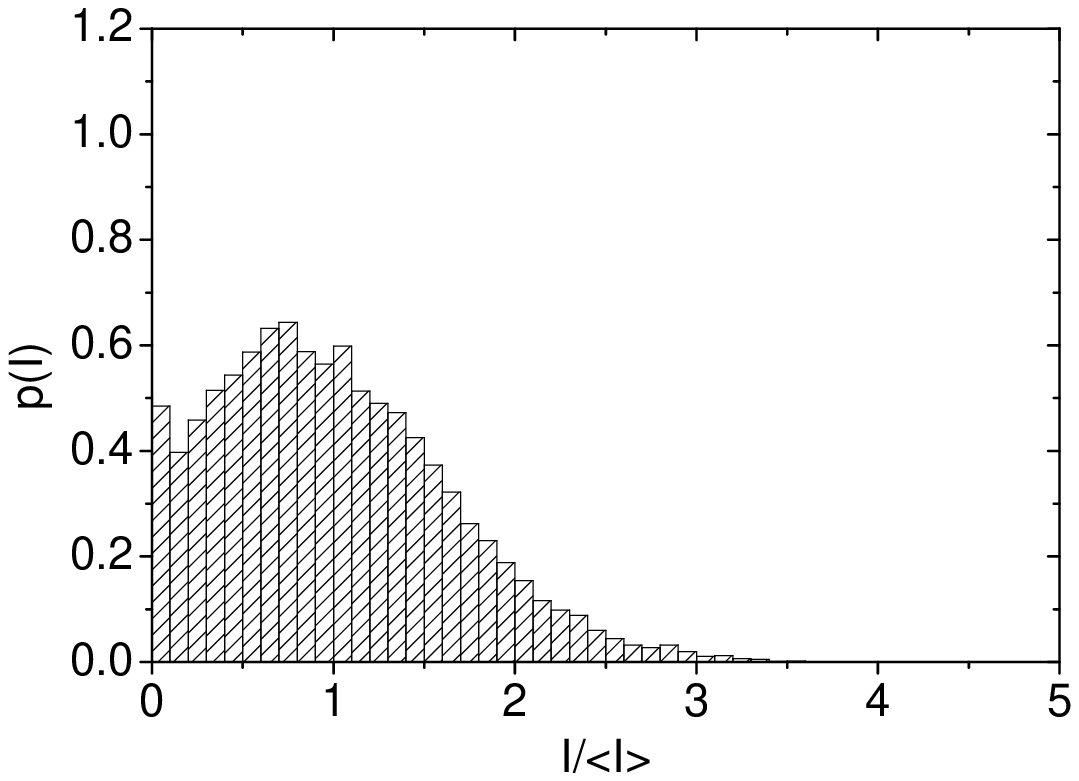}
\includegraphics[width=0.32\textwidth]{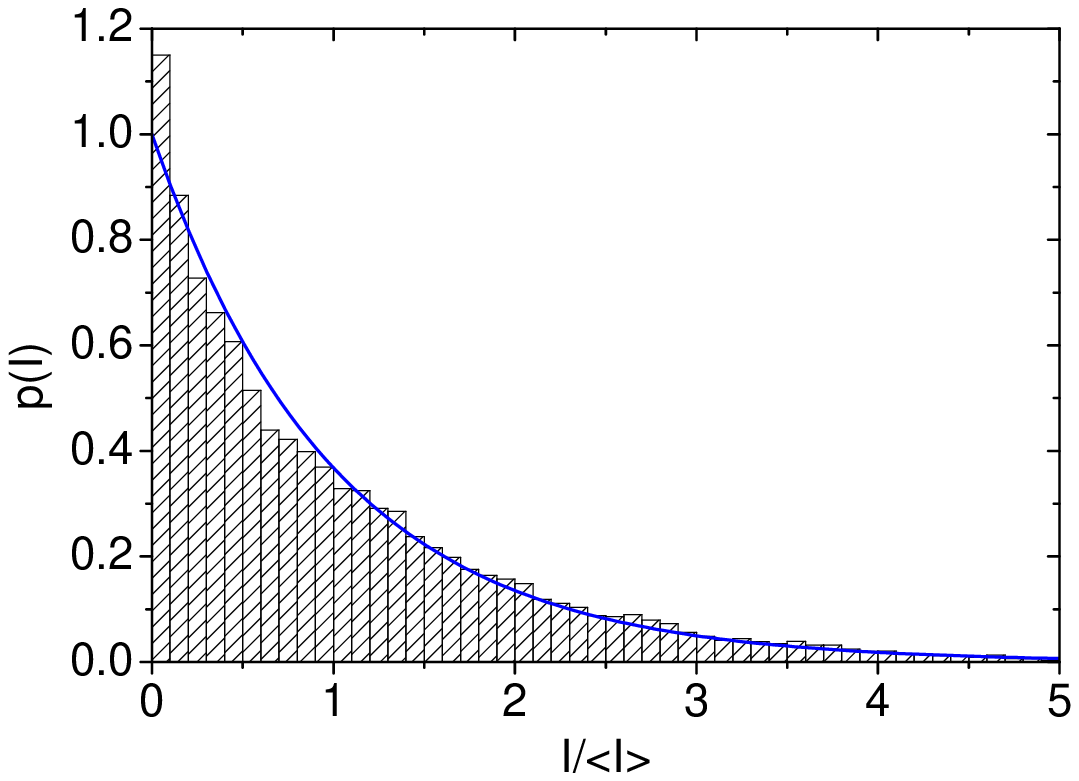}

\includegraphics[width=0.32\textwidth]{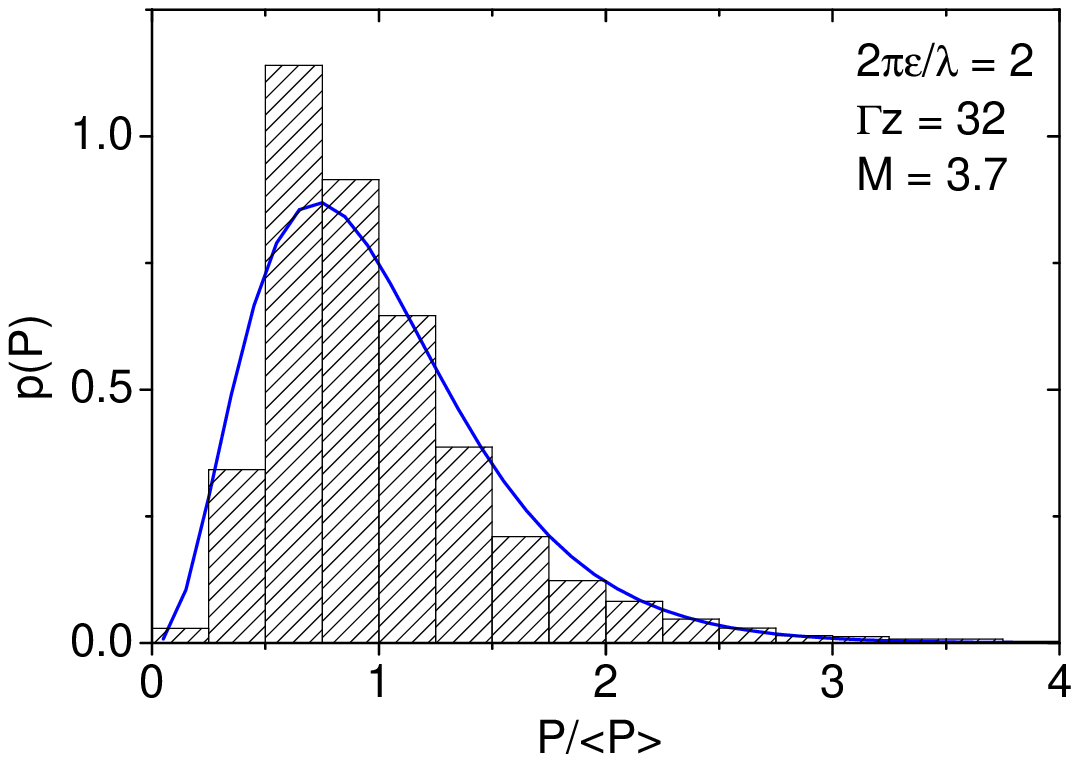}
\includegraphics[width=0.32\textwidth]{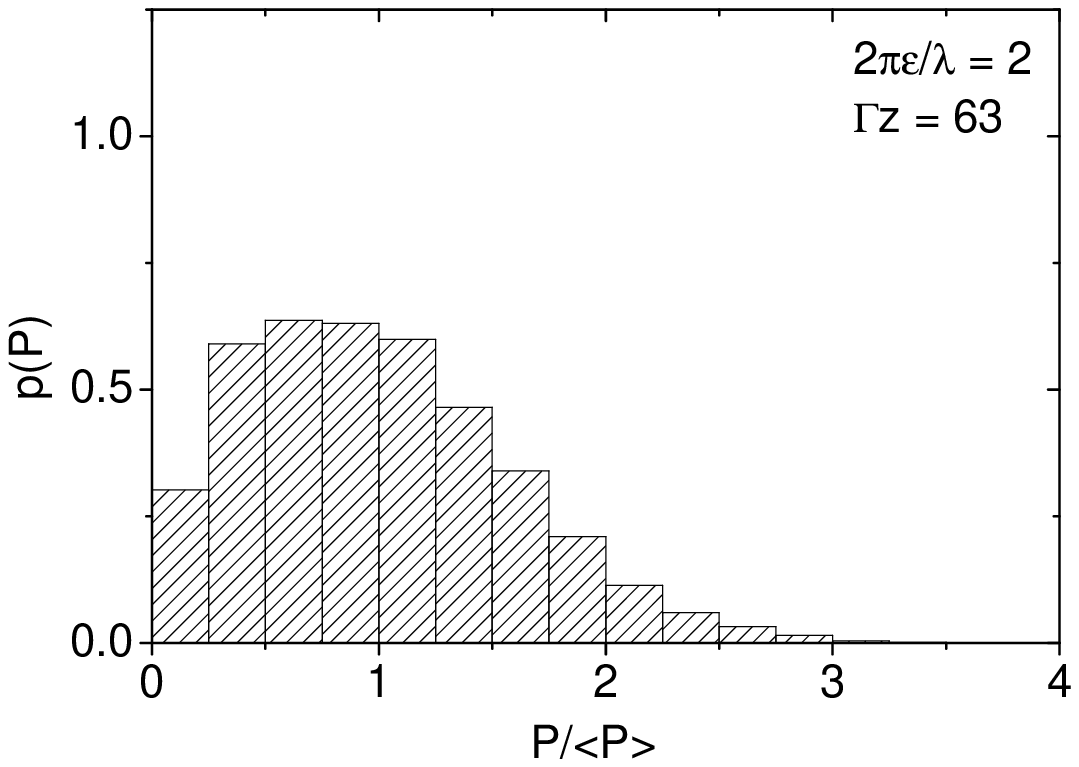}
\includegraphics[width=0.32\textwidth]{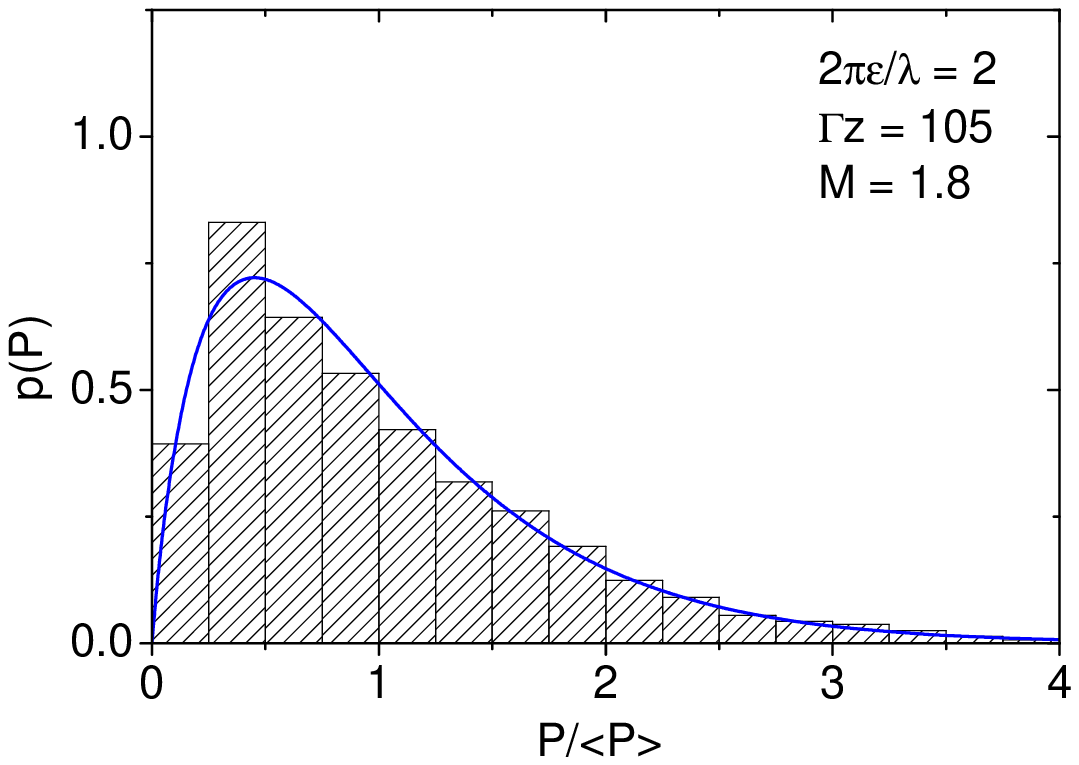}

\caption{
Probability density distributions of the instantaneous power density
$I = |\tilde{E}|^2$ (top), and of the instantaneous radiation power $P$
(bottom)
from
SASE FEL for different stages of amplification:
linear regime ($\hat{z} = 32)$,
saturation regime ($\hat{z} = 63)$,
and deep nonlinear regime ($\hat{z} = 105$).
Solid lines on the power density histograms (top)
 represent negative exponential distribution
(\ref{neg-exp-1}).
Solid lines on power histograms (bottom)
 represent gamma distribution
(\ref{gamma})
 with $M = 1/\sigma_{\mathrm{P}}^2$
(see Fig.~\ref{fig:pzem2}).
Here $\hat{\epsilon } = 2$.
}
\label{fig:hprobem2}
\end{figure}

\begin{equation}
p({P}) = \frac{M^M}{\Gamma (M)}
\left( \frac{P}{\langle P \rangle }\right)^{M-1}
\frac{1}{\langle P \rangle } \exp \left( -M \frac{P}
{\langle P \rangle } \right) \ ,
\label{gamma}
\end{equation}

\noindent where $\Gamma (M)$ is the gamma function with argument $M$,
and

\begin{equation}
M = \frac{1}{\sigma_{\mathrm{P}}^2} \ ,
\label{m-parameter}
\end{equation}

\noindent and $\sigma_{\mathrm{P}}^2 = \langle (P -\langle P \rangle )^2 \rangle /
\langle P \rangle^2 $ is the relative dispersion of the radiation power. Note
that for completely chaotic polarized light parameter $M$ has clear physical
interpretation, it is the number of modes \cite{book}. Thus, it becomes clear
that the relative dispersion of the radiation power directly relates to the
coherence properties of the SASE FEL operating in the linear regime. The degree
of transverse coherence in this case can be naturally defined as:

\begin{equation}
\zeta  = \frac{1}{M} = \sigma^2_{\mathrm{P}} \ .
\label{eq:degcoh-m}
\end{equation}

\noindent Indeed, in the linear regime we deal with a Gaussian random
process, and the power density fluctuates in accordance with the negative
exponential distribution and its relative width is equal to 1. If there
is full transverse coherence then the same refers to the power. If the
radiation is partially coherent, then we have a more general law for
power fluctuations, namely the gamma distribution (\ref{gamma}). In the linear
regime fluctuations of the radiation power, $\sigma _{\mathrm{P}}$, grow
steadily with the undulator length (see Fig.~\ref{fig:pzem2}) because of
mode selection process (\ref{eq:modes}).

\begin{figure}[b]
\includegraphics[width=0.5\textwidth]{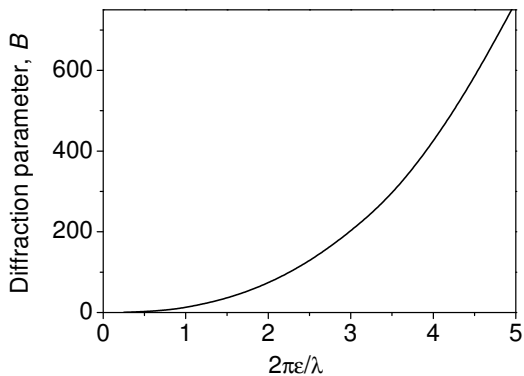}
\caption{
Diffraction parameter $B$ versus parameter $\hat{\epsilon } $
for optimized XFEL.
}
\label{fig:difparopt}
\end{figure}

Another physical parameter of the problem relating to the transverse
coherence is diffraction parameter $B$.
Mechanism of formation of transverse coherence is rather transparent. If
diffraction expansion of the radiation on a scale of the field gain length is
comparable with the transverse size of the electron beam, we can expect a high
degree of transverse coherence. For this range of parameters the value of the
diffraction parameter is small. If diffraction expansion of the radiation is
small (which happens at large values of the diffraction parameter) then we can
expect significant degradation of the degree of transverse coherence. This
happens simply because different parts of the beam produce radiation nearly
independently. In terms of the radiation expansion in the eigenmodes
(\ref{eq:modes}) this range of parameters corresponds to degeneration
of modes (see Appendix A). Diffraction parameter for optimized XFEL exhibits
strong dependence on the parameter $\hat{\epsilon }$
(see eq.~(\ref{eq:bopt}) and Fig.~\ref{fig:difparopt}), and we can expect
that the degree of transverse
coherence should drop rapidly with the increase of the parameter $\hat{\epsilon
}$.

An important physical quantity describing random fields is the coherence time.
Figure~\ref{fig:slice-along-z-em2} gives qualitative picture of formation of
longitudinal coherence in SASE FEL. At the beginning of the undulator
the radiation is simply incoherent undulator radiation. When amplification
process starts to dominate over spontaneous emission, we obtain formation of
spikes (wavepackets). The width of the spikes defines the coherence time. In
the high gain linear regime the width of spikes grows with the undulator
length, and the coherence time also grows proportionally to the square root of
the undulator length. It achieves maximum value in the end of linear regime
and then decreases rapidly in the nonlinear regime (see
Fig.~\ref{fig:tcem2}). The maximum value of the coherence time depends on
the saturation length and, therefore, on the value of the parameter
$N_{\mathrm{c}}$ \cite{statistics-oc}.

We see that physical background defining general features of the radiation
from SASE FEL operating in the linear regime is rather transparent. Despite the
behavior of the SASE FEL in the nonlinear regime is rather complicated (see
Figs.~\ref{fig:pzem2} and \ref{fig:tcem2}), we will show below that the main
characteristics of the SASE FEL operating at the saturation point have rather
simple physical scaling.

\section{General definitions}

\subsection{Degree of transverse coherence}

    The transverse coherence properties of the radiation are described in terms
of the transverse correlation functions.  The first-order transverse
correlation function is defined as

\begin{displaymath}
\gamma_1 (\vec{r}_{\perp}, \vec{r}\prime _{\perp}, z, t) = \frac{
\langle \tilde{E} (\vec{r}_{\perp}, z, t)
\tilde{E}^{*} (\vec{r}\prime _{\perp}, z, t) \rangle }
{ \left[ \langle |\tilde{E} (\vec{r}_{\perp}, z, t) |^2 \rangle
\langle |\tilde{E} (\vec{r}\prime _{\perp}, z, t) |^2 \rangle \right]^{1/2}}
\ ,
\end{displaymath}

\noindent where $\tilde{E}$ is the slowly varying amplitude of the
amplified wave:

\begin{equation}
E =
\tilde{E} (\vec{r}_{\perp}, z, t) \E^{\I \omega_0 (z/c - t)}
+ {\mathrm{C.C.}} \ .
\label{eq:space-corr}
\end{equation}

\noindent In the following we consider the model of a stationary random
process, thus assuming that $\gamma_1$ does not depend on time. We define the
degree of transverse coherence as:

\begin{equation}
\zeta =
\frac
{
\int |\gamma _1 (\vec{r}_{\perp},\vec{r}\prime _{\perp})|^{2}
I(\vec{r}_{\perp})
I(\vec{r}\prime _{\perp})
\D\vec{r}_{\perp}
\D\vec{r}\prime _{\perp}
}
{
[\int I(\vec{r}_{\perp}) \D\vec{r}_{\perp}]^{2}
} \ .
\label{eq:def-degcoh}
\end{equation}

\noindent When SASE FEL operates in the linear regime such a definition for the
degree of transverse coherence is mathematically equivalent to
(\ref{eq:degcoh-m}) expressed in terms of the relative dispersion of the
instantaneous radiation power, $\sigma_{\mathrm{P}}^2$. Analysis of the
asymptotic of the deep nonlinear regime (see Fig.~\ref{fig:tcem2}) shows that
surprisingly the degree of transverse coherence defined by
(\ref{eq:def-degcoh}) again tends to be an agreement with (\ref{eq:degcoh-m}).
This feature has deep physical background. When amplification process just
enters nonlinear stage, the statistics of the radiation is not Gaussian
anymore. In particular,  the probability distribution function of the radiation
power density, is not negative exponential distribution (see
Fig.~\ref{fig:hprobem2}). Thus, definition of the degree of transverse coherence
(\ref{eq:degcoh-m}) has no physical sense near the saturation point. However,
simulations show that in the deep nonlinear regime the probability distribution
of the radiation power density again tends to the negative exponential
distribution. This gives us a hint that the properties of the radiation from
SASE FEL operating in the deep nonlinear regime tend again to be the properties
of completely chaotic polarized light. Similar asymptotical behavior has been
observed in the framework of the one-dimensional model as well
\cite{statistics-oc}.

\subsection{Coherence time}

   Longitudinal coherence is described in terms of time correlation functions.
The first order time correlation function, $g_1(t,t')$, is calculated in
accordance with the definition:

\begin{equation}
g_1(\vec{r},t-t')  =
\frac{\langle \tilde{E}(\vec{r},t)\tilde{E}^*(\vec{r},t')\rangle }
{\left[\langle \mid\tilde{E}(\vec{r},t)\mid^2\rangle
\langle \mid\tilde{E}(\vec{r},t')\mid^2\rangle \right]^{1/2}} \ ,
\label{def-corfunction}
\end{equation}

\noindent For a stationary random process time correlation functions are
functions of the only argument, $\tau = t - t'$. The coherence time is
defined as \cite{book}:

\begin{equation}
\tau_{\mathrm{c}} = \int \limits^{\infty}_{-\infty}
| g_1(\tau) |^2 \D\tau \ .
\label{coherence-time-def}
\end{equation}

\subsection{Brilliance and degeneracy parameter}

Main figure of merit of the SASE FEL performance is brilliance, i.e. density of
photons in the phase space. In fact, the brilliance is proportional to the
degeneracy parameter $\delta $, i.e. the number of photons per mode (coherent state).
Note that when $\delta \gg 1$, the classical statistics is applicable, while quantum
description of the field is necessary as soon as $\delta$ is comparable to (or less than) one.
Using the definitions of the degree of transverse coherence (\ref{eq:def-degcoh}) and
coherence time (\ref{coherence-time-def}), one can define degeneracy parameter:

\begin{equation}
\delta = \dot{N}_{ph} \tau_{\mathrm{c}} \zeta \ ,
\label{degeneracy-def}
\end{equation}

\noindent where $\dot{N}_{ph} = N_{ph}^{tot}/T$ is the photon flux,
$N_{ph}^{tot}$ is the total number of photons in a long flat-top pulse of
duration $T$ (as everywhere in this paper we consider ensemble average values).
The definition (\ref{degeneracy-def}) is perfect for a Gaussian random process
(that we have in linear regime). Indeed, degree of transverse coherence is an
inverse number of transverse modes (\ref{eq:degcoh-m}), while
$\tau_{\mathrm{c}}/T$ is an inverse number of longitudinal modes within the
pulse \cite{book,goodman}. Thus, degeneracy parameter is equal to the number of
photons per pulse divided by total number of modes per pulse (that is equal to
the squared inverse rms fluctuations of pulse energy \cite{book}). It can be
directly measured in an experiment for SASE FEL operating in the linear regime.
We use (\ref{degeneracy-def}) as a natural generalization to
characterize SASE FEL properties at saturation.

Let us  introduce a notion of
normalized degeneracy parameter

\begin{equation}
\hat{\delta} =
\hat{\eta} \hat{\tau}_{\mathrm{c}} \zeta \ .
\label{eq:norm-deg-par}
\end{equation}

\noindent Here normalized FEL efficiency is defined as $\hat{\eta} = P /(\rho
P_{\mathrm{b}})$ where $P$ is radiation power, and
$P_{\mathrm{b}} = \gamma mc^2I/e$ is electron beam power.
Normalized coherence time is defined as
$\hat{\tau}_{\mathrm{c}} = \rho \omega \tau_{\mathrm{c}}$. Parameter
$\hat{\delta }$ and the degeneracy parameter $\delta $ are simply related as:

\begin{equation}
\delta = \frac{P_{\mathrm{b}}}{\hbar \omega ^2} \hat{\delta } \ ,
\label{eq:deg-par}
\end{equation}

\noindent or, in practical units

\begin{equation}
\delta \simeq
2.7\times 10^7 \times
\lambda^2 [\mathrm{\AA }\ ] \times
I [\mathrm{kA}] \times
E [\mathrm{GeV}] \times \hat{\delta } \ ,
\label{eq:deg-par-pract}
\end{equation}

\noindent where $E = \gamma mc^2$ is the electron energy.
Note that the degeneracy parameter is very large even for a SASE FEL operating
at the wavelength of 0.1~nm. With multi-kA electron beams and other relevant
parameters, mentioned in Table 1, degeneracy parameter would be of the order of
$10^8$ - $10^9$. Thus, a classical treatment of SASE FEL is justified.

Let us now turn to the calculation of peak brilliance. It is defined as a transversely
coherent spectral flux:

\begin{equation}
B_r =
\frac{\omega \D \dot{N}_{ph}}{\D \omega} \
\frac{\zeta}{\left(\frac{\lambda}{2}\right)^2} \ .
\label{eq:bril-1}
\end{equation}

\noindent The spectrum of SASE FEL radiation in a high-gain linear regime has a Gaussian shape,
it is also close to the Gaussian at saturation point \cite{book}. In this case

\begin{displaymath}
\frac{\omega \D \dot{N}_{ph}}{\D \omega} =
\frac{\omega \dot{N}_{ph}}{\sqrt{2\pi} \sigma_{\omega}} \ ,
\end{displaymath}

\noindent where $\sigma_{\omega}$ is the rms bandwidth. For a Gaussian line, with the definition of
coherence time (\ref{coherence-time-def}), one gets \cite{book}:

\begin{displaymath}
\tau_{\mathrm{c}} = \frac{\sqrt{\pi}}{\sigma_{\omega}} \ .
\end{displaymath}

\noindent Thus, it follows from (\ref{eq:norm-deg-par}) and (\ref{eq:bril-1}) that

\begin{equation}
B_r = \frac{4\sqrt{2} c}{\lambda^3} \ \delta \ .
\label{eq:bril-2}
\end{equation}

\noindent The peak brilliance can then be calculated as follows

\begin{equation}
B_r [\mathrm{ phot./(sec. \ mrad^2 \ mm^2 \ 0.1\% \ bandw.)}] \simeq
4.5\times 10^{31} \times
\frac{I [\mathrm{kA}] \times
E [\mathrm{GeV}]}{\lambda [\mathrm{\AA } \ ]}
\times \hat{\delta } \ .
\label{eq:bril-pract}
\end{equation}

\noindent For future SASE FELs, operating at 1~\AA \ , an expected value of the
peak brilliance is $10^{32}$-$10^{33}$.

An important feature of our analysis is application of similarity
techniques to the analysis of simulation results which allows to derive
universal parametric dependencies of the output characteristics of the
radiation. As we mentioned above, within accepted approximations (optimized
SASE FEL and negligibly small energy spread in the electron beam),  output
characteristics of SASE FEL are universal functions of two parameters,
normalized undulator length $\hat{z} = \Gamma z$ and parameter
$\hat{\epsilon}$. If one traces evolution of the brilliance (or, degeneracy
parameter) of the radiation along the undulator length  there is always the
point, which we define as the saturation point, where the brilliance reaches
maximum value (see Fig.~\ref{fig:tcem2}). In the following we present
characteristics of the radiation at the saturation point which are universal
functions of the only parameter $\hat{\epsilon}$.

\section{Properties of the radiation from optimized XFEL operating in
the saturation regime}

\begin{figure}[b]

\includegraphics[width=0.32\textwidth]{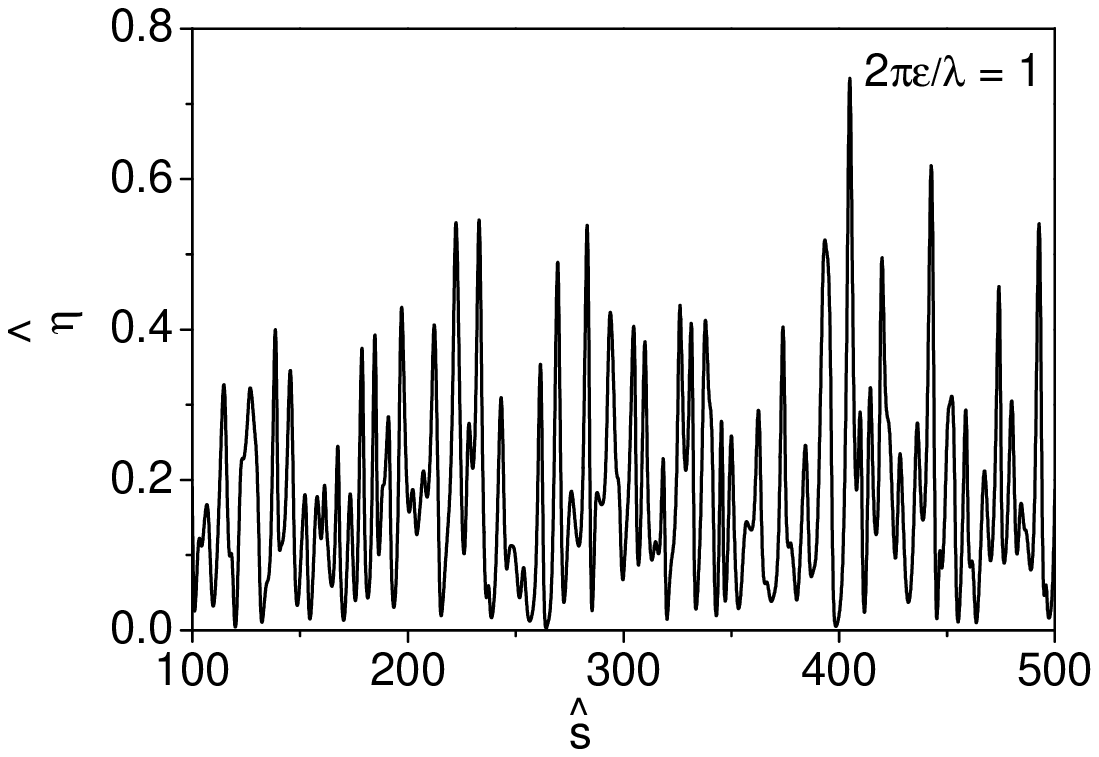}
\includegraphics[width=0.32\textwidth]{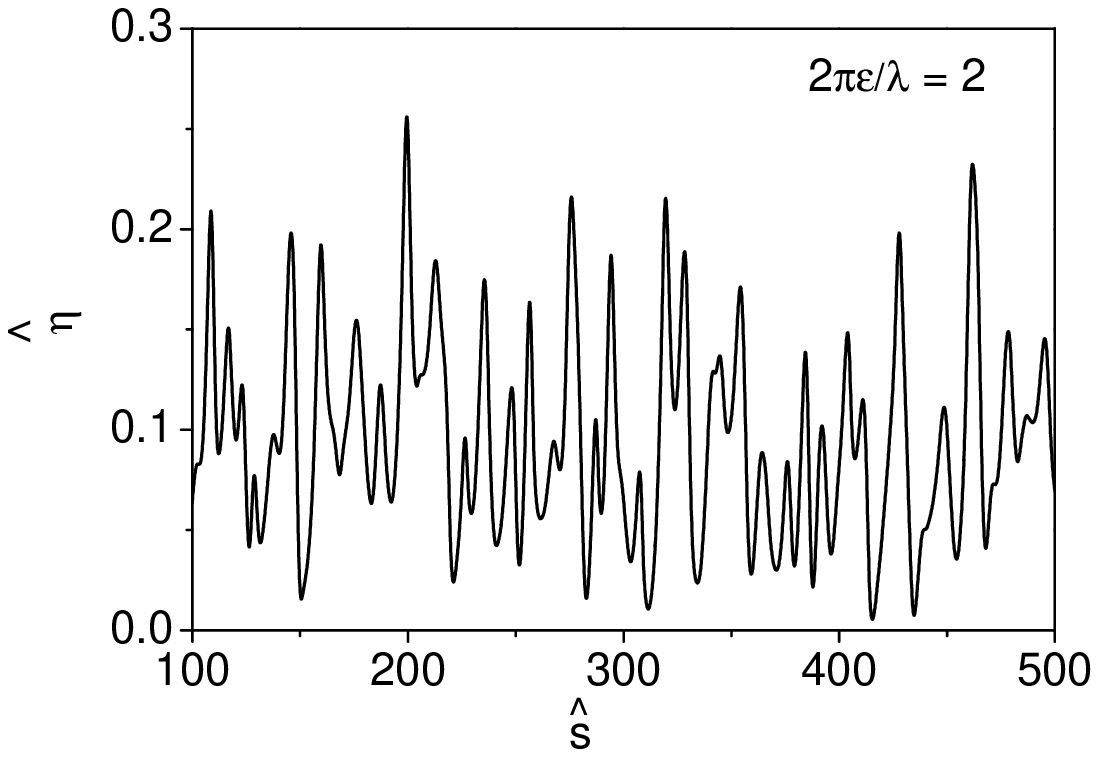}
\includegraphics[width=0.32\textwidth]{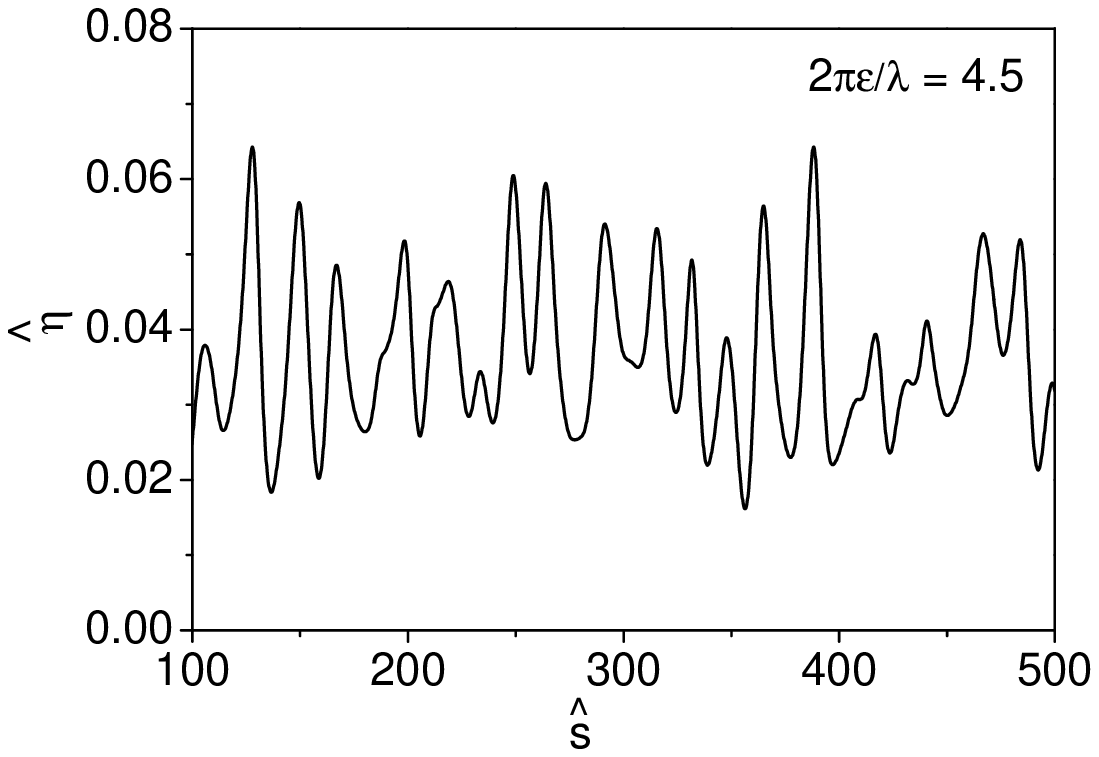}

\hspace*{0.05\textwidth}
\includegraphics[width=0.23\textwidth]{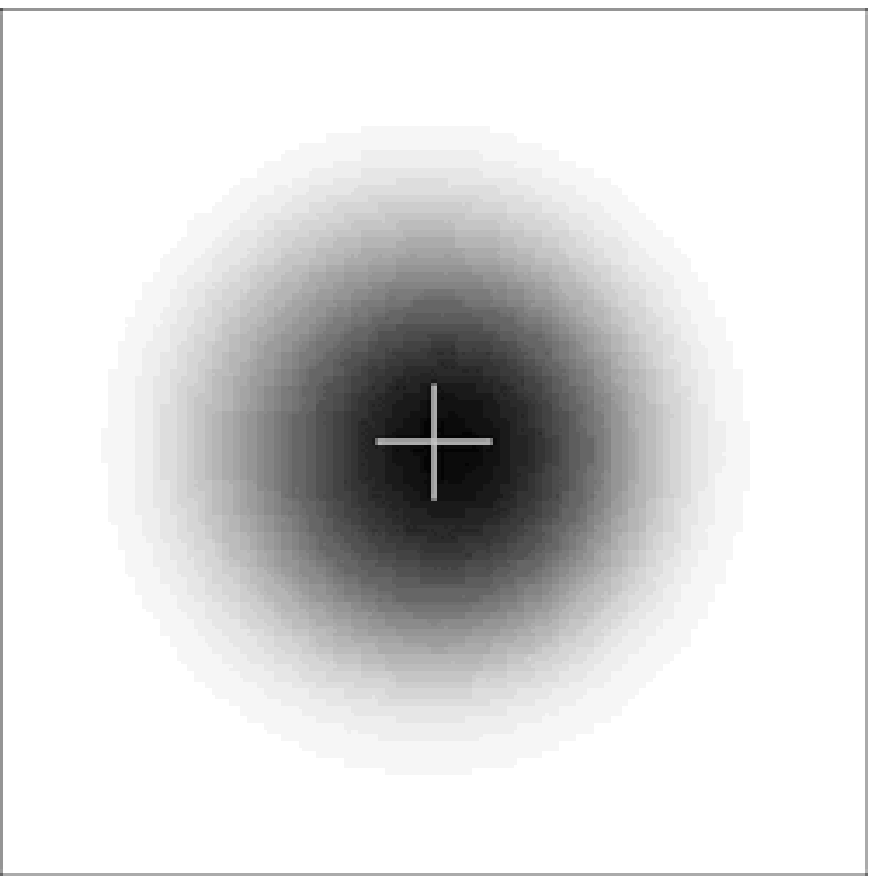}
\hspace*{0.1\textwidth}
\includegraphics[width=0.23\textwidth]{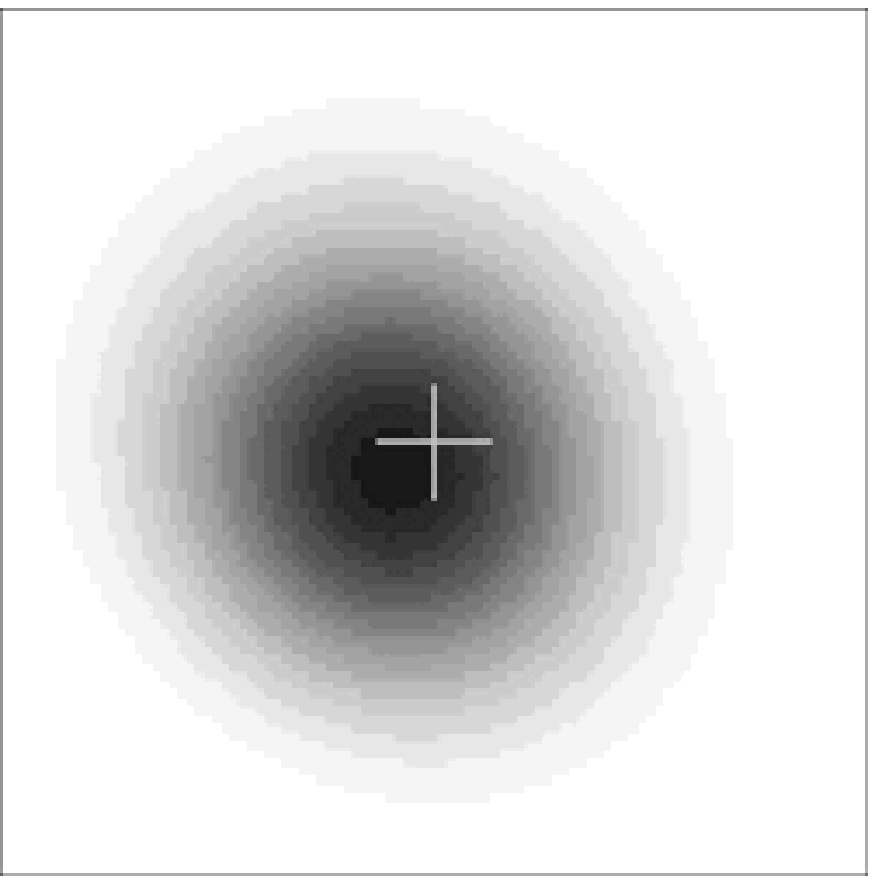}
\hspace*{0.1\textwidth}
\includegraphics[width=0.23\textwidth]{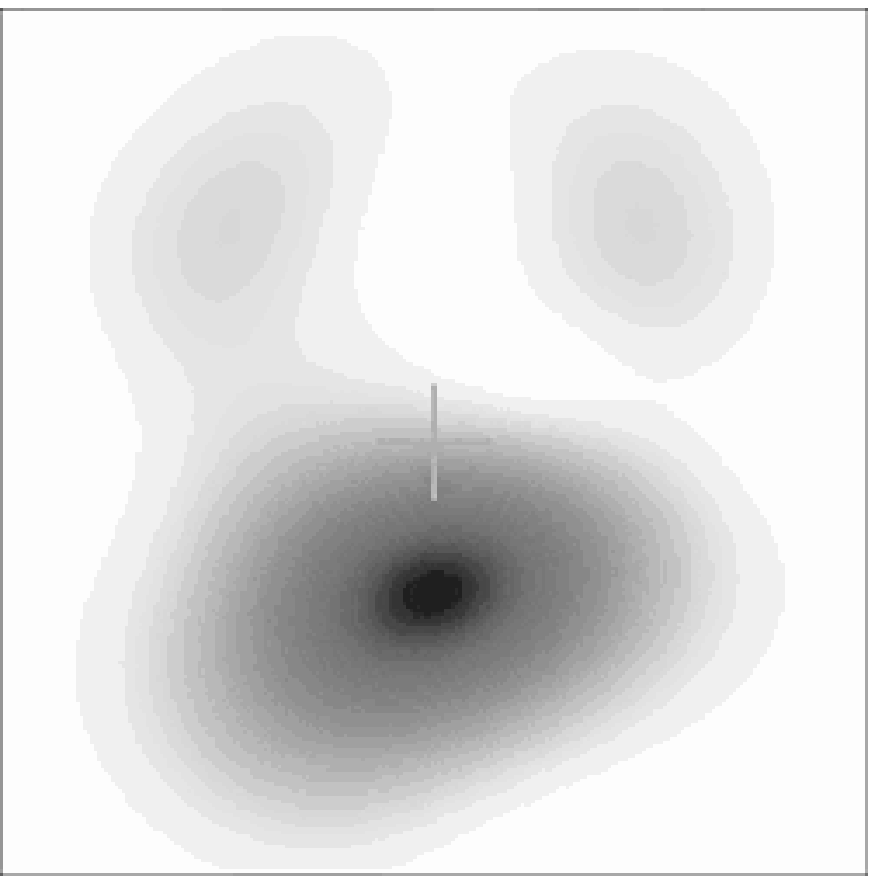}

\caption{Top:
normalized power in a radiation
pulse, $\hat{\eta }$ versus
$\hat{s} = \rho \omega (z/\bar{v}_{z}-t)$.
Bottom: typical power density distribution in a slice of the radiation.
SASE FEL operates in the saturation.
Values of $\hat{\epsilon }  = 1$, 2 and 4.5
correspond to the left, middle, and right plot,
respectively. Crosses show geometrical center of the beam.
}

\label{fig:psattemporal}
\end{figure}

\begin{figure}[tb]
\includegraphics[width=0.5\textwidth]{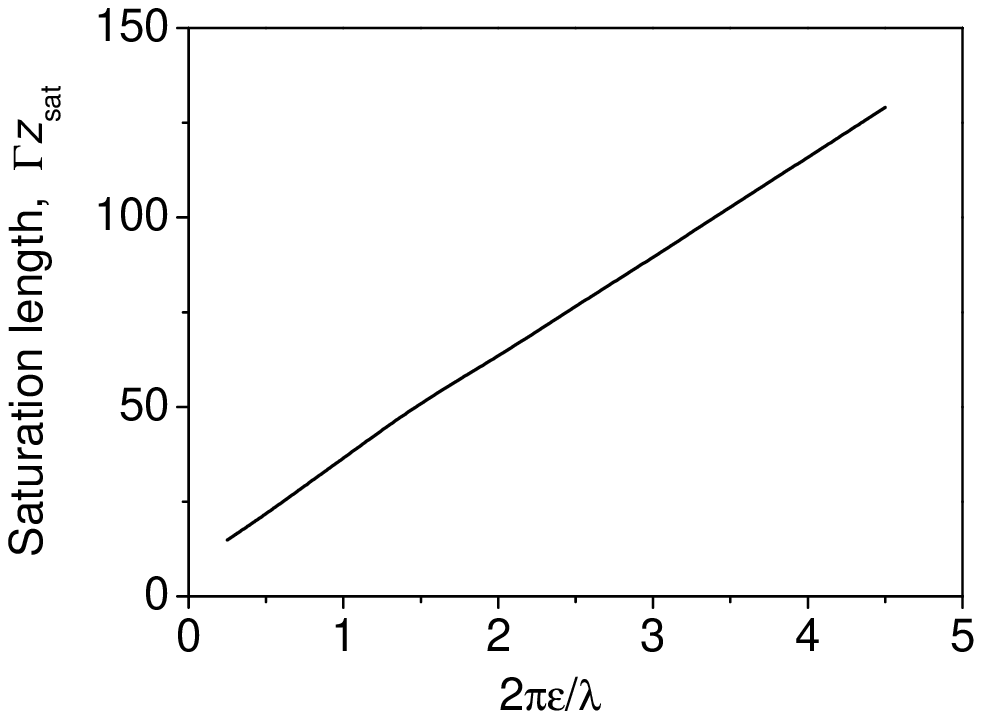}
\caption{
Saturation length $\hat{z}_{\mathrm{sat}} = \Gamma z_{\mathrm{sat}}$
versus parameter $\hat{\epsilon } $.
}
\label{fig:zsatopt}

\includegraphics[width=0.49\textwidth]{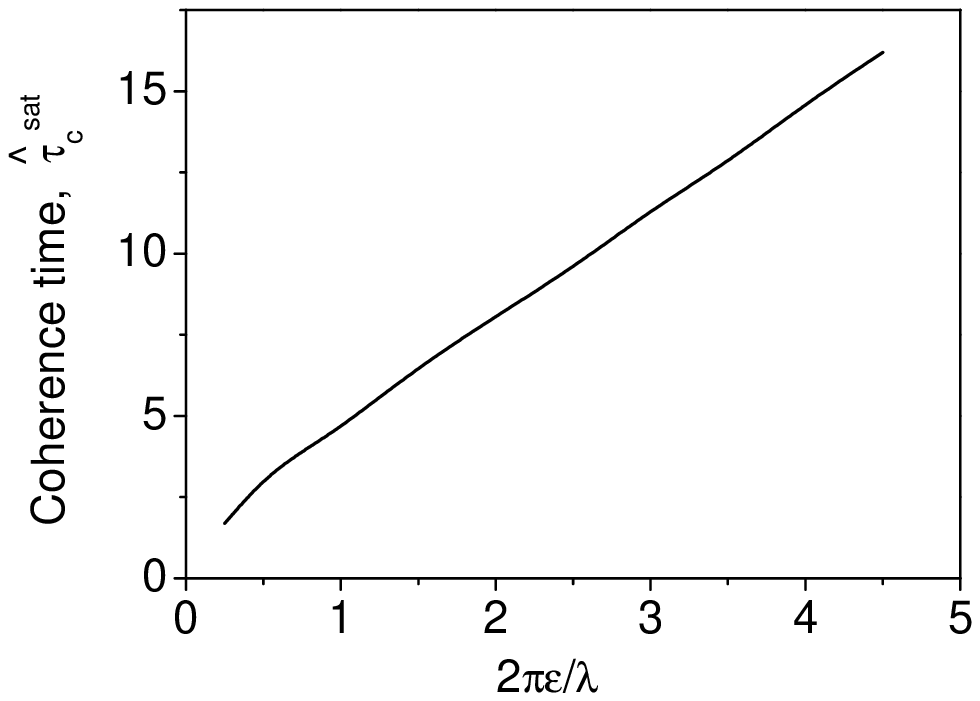}
\includegraphics[width=0.49\textwidth]{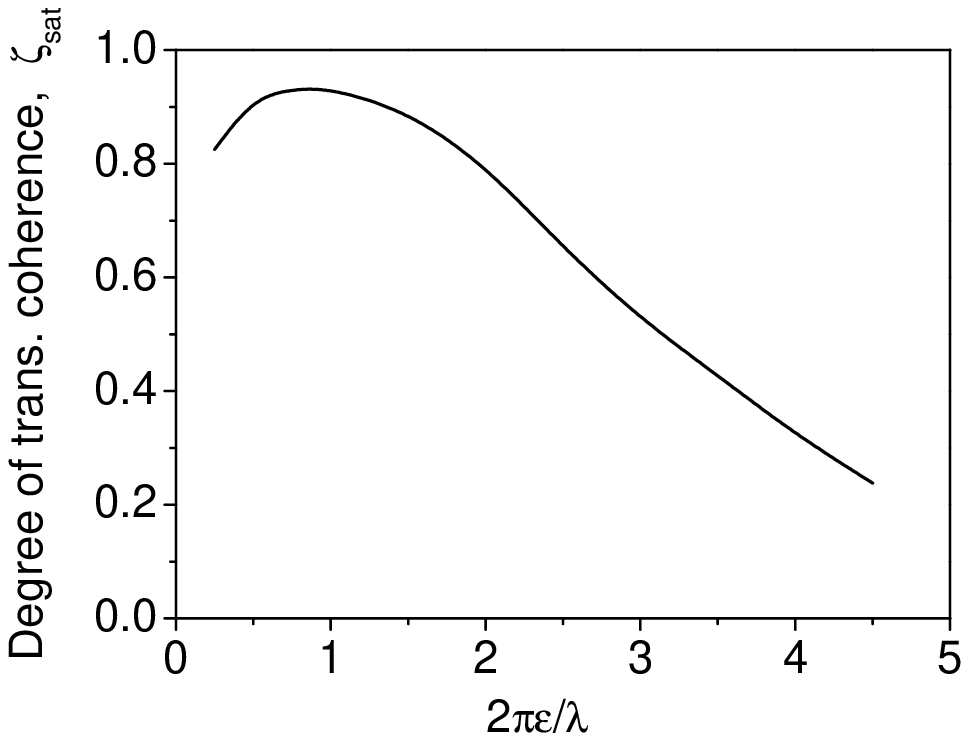}

\caption{
Degree of transverse coherence, $\zeta _{\mathrm{sat}}$, and normalized
coherence time, $\hat{\tau }_{\mathrm{c}}^{\mathrm{sat}}$ in the saturation
versus parameter $\hat{\epsilon } $.
}
\label{fig:dcohsat}

\includegraphics[width=0.49\textwidth]{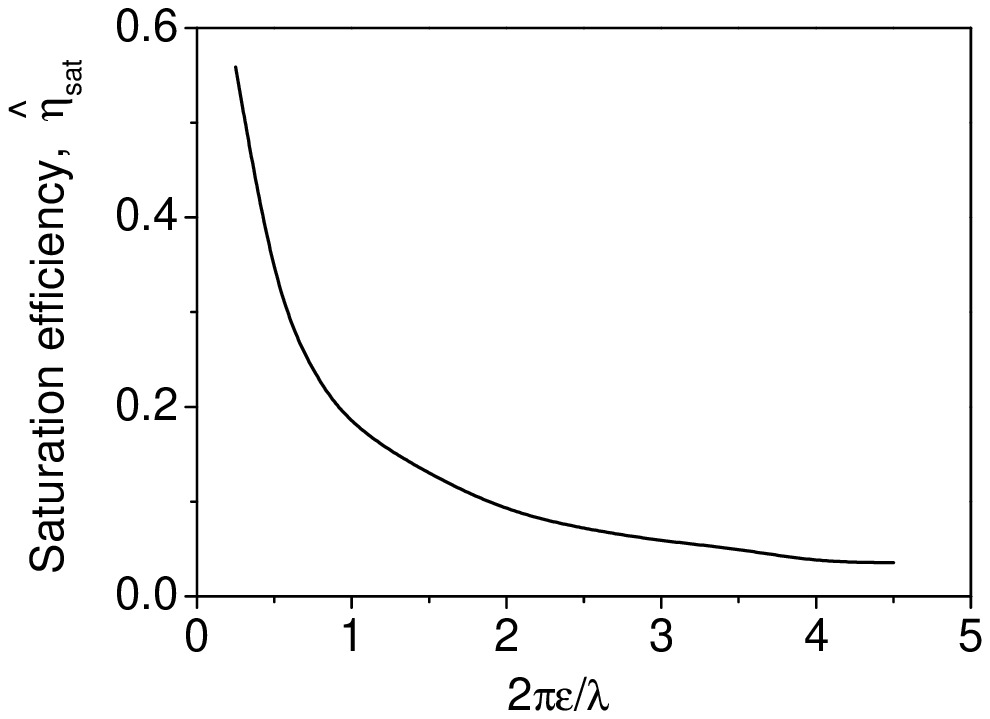}
\includegraphics[width=0.49\textwidth]{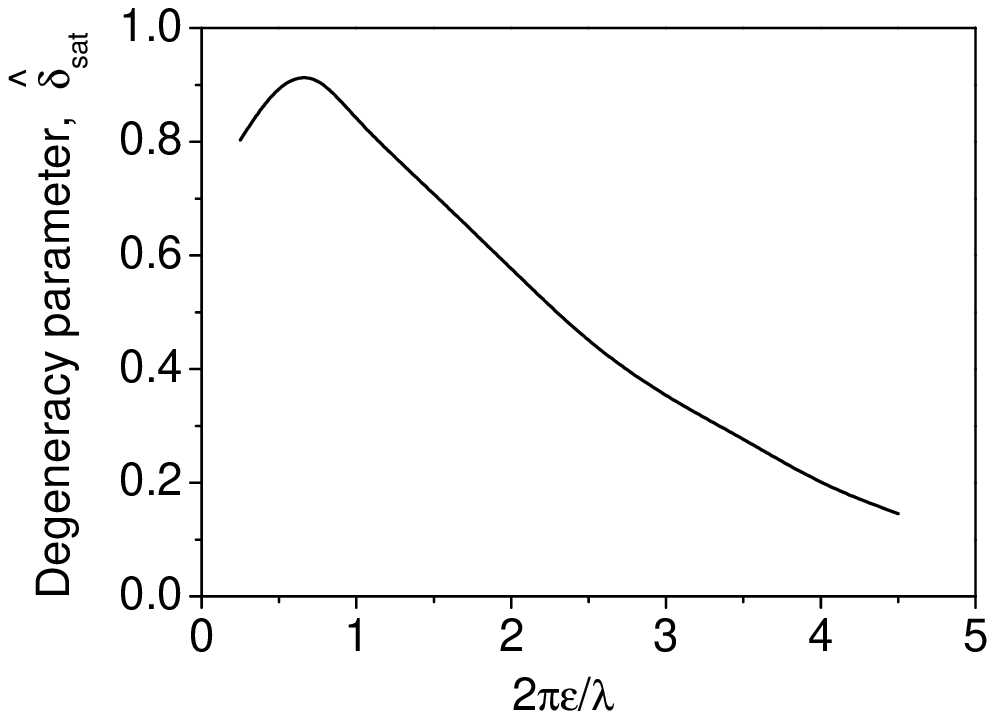}

\caption{
Averaged efficiency, $ \langle \hat{\eta }_{\mathrm{sat}} \rangle $,
and normalized degeneracy parameter,
$\hat{\delta}_{\mathrm{sat}}$, in the saturation
versus parameter $\hat{\epsilon }$.
}
\label{fig:brilsat}
\end{figure}

Simulations of the FEL process have been performed for the case of a long bunch
with uniform axial profile of the beam current. Such a model provides rather
accurate predictions for the coherence properties of the XFEL, since typical
radiation pulse from the XFEL is much longer than the coherence time.
Calculations has been performed with FEL simulation code FAST using actual
number of electrons in the beam. The value of parameter
$N_{\mathrm{c}} = 8 \times 10^{5}$ corresponds to the parameter range of XFEL
operating at the radiation wavelength about 0.1~nm.

Figure~\ref{fig:psattemporal} gives visual picture of the slice properties of
the radiation at the saturation for different values of the parameter
$\hat{\epsilon }$. Saturation point is defined as the point where the radiation
achieves maximum brilliance (or, maximum degeneracy
parameter (\ref{eq:norm-deg-par})). A series of simulation runs has been
performed in the range of the parameter $\hat{\epsilon } = 0.25 \ldots 4.5$.
Application of similarity techniques described above allowed us to extract
universal parametric dependencies of the main characteristics of the optimized
XFEL operating in the saturation regime (see
Figs.~\ref{fig:zsatopt}-\ref{fig:brilsat}).

Figure~\ref{fig:zsatopt} shows the dependence of the saturation length
$\hat{z}_{\mathrm{sat}} = \Gamma z_{\mathrm{sat}}$ on parameter $\hat{\epsilon
}$. Analysis of the curve shows that the saturation length scales as
$\hat{z}_{\mathrm{sat}} \propto \hat{\epsilon }^{5/6}$. Such dependence
directly follows from the optimization procedure of the gain length given by
(\ref{eq:lg}).
The normalized coherence time in the saturation regime,
$\hat{\tau}_{\mathrm{c}}^{\mathrm{sat}}$ is also proportional to $\hat{\epsilon
}^{5/6}$ (see Fig.~\ref{fig:dcohsat}).

\begin{figure}[tb]
\includegraphics[width=0.5\textwidth]{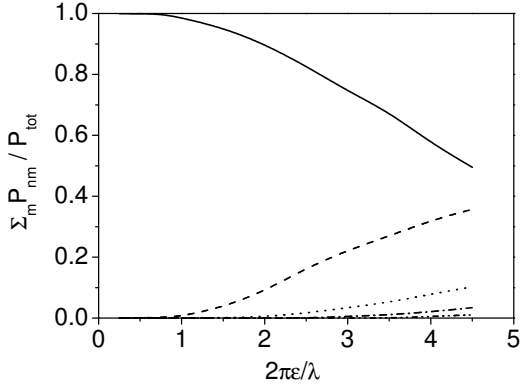}
\caption{
Partial contributions of the modes with azimuthal index $m = 0 \ldots 4$ into
the total power
versus parameter $\hat{\epsilon }$. SASE FEL operates in the saturation.
}
\label{fig:partpow}
\end{figure}

The dependence of the degree of transverse coherence in the saturation regime
on the parameter $\hat{\epsilon }$ exhibits rather complicated behavior (see
Fig.~\ref{fig:dcohsat}). It reaches maximum
value in the range of $\hat{\epsilon }$ values about of unity, and drops at
small and large values of $\hat{\epsilon }$. Actually, the degree of transverse
coherence is formed due to two effects. The first effect takes place due to
interdependence of the poor longitudinal coherence and transverse coherence
\cite{trcoh-oc}. Due to the start-up from shot noise every radiation mode
entering (\ref{eq:modes}) is excited within finite spectral bandwidth.
This means that in the high gain linear regime the radiation of the SASE FEL is
formed by many fundamental TEM$_{00}$ modes with different frequencies.  The
transverse distribution of the radiation field of the mode is also different
for different frequencies. Smaller value of the diffraction parameter (i.e.
smaller value of $\hat{\epsilon }$) corresponds to larger deviation of the
radiation mode from the plane wave. This explains a decrease of the transverse
coherence at small values of $\hat{\epsilon }$. When the parameter
$\hat{\epsilon }$ increases, the diffraction parameter increases as well thus
leading to the degeneration of the radiation modes. Amplification process in the
SASE FEL passes limited number of the field gain lengths, and
starting from some value of $\hat{\epsilon }$ the linear stage of amplification
becomes too short to provide mode selection process (\ref{eq:modes}). When
amplification process enters nonlinear stage, the mode content of the radiation
becomes even more rich due to independent growth of the radiation modes in the
nonlinear medium (see Fig.~\ref{fig:partpow}). Thus, at large values of
$\hat{\epsilon }$ the degree of transverse coherence is limited by poor mode
selection. Analytical estimations, presented in Appendix A, show that in the
limit of large emittance, $\hat{\epsilon } \gg 1$, the degree of transverse
coherence scales as $1/\hat{\epsilon }^2$.

We present in Fig.~\ref{fig:brilsat} the plots for normalized
efficiency and degeneracy parameter for optimized XFEL. Normalized efficiency
in saturation has simple scaling, it falls inversely proportional to the
parameter $\hat{\epsilon }$. Taking into account that the value of the
coherence time $\hat{\tau}_{\mathrm{c}}^{\mathrm{sat}}$ scales proportional to
$\hat{\epsilon }^{5/6}$, we find that the normalized degeneracy parameter of
the radiation is nearly proportional to the degree of transverse coherence,
$\hat{\delta }_{\mathrm{sat}} \propto \zeta / \hat{\epsilon}^{1/6}$.

\begin{table}[tb]
\caption{Parameter space of XFEL projects}
\begin{tabular}{ l l l l l }
\hline
                          & \multicolumn{2}{c}{European  XFEL}        & LCLS
& SCSS \\
                          & SASE1      & SASE2            &          & \\
\hline
Radiation wavelength, nm  & 0.1       & 0.15            & 0.15     &  0.1 \\
Beam energy, GeV          & 17.5      & 17.5            & 14.35    & 6.135 \\
rms normalized emittance $\epsilon _n$, mm-mrad
			  & 1.4       & 1.4             & 1.2      & 0.85 \\
Parameter $\hat{\epsilon } $
			  & 2.6       & 1.7             & 1.8      & 4.5 \\
Degree of transverse coherence $\zeta $
			  & 0.65      & 0.85            & 0.83     & 0.24 \\
\hline
\end{tabular}
\label{tab:xfel-par}
\end{table}

Finally, in Table~\ref{tab:xfel-par} we present comparison of existing XFEL
projects, the European XFEL, LCLS and SCSS in terms of degree of transverse
coherence \cite{euro-xfel-tdr,lcls-cdr,scss-cdr}. We see that the European XFEL
and LCLS are in the same range of parameter space. These projects assume
conservative value of the emittance, and relatively high degree of transverse
coherence is achieved by increasing the energy of the driving accelerator.
Project SCSS assumes much smaller energy of the driving accelerator. Thus,
despite much smaller value of the normalized emittance it falls in the range of
parameters for the output radiation with poor transverse coherence.

\clearpage

\appendix

\renewcommand{\theequation}{\Alph{section}.\arabic{equation}}

\setcounter{equation}{0}
\setcounter{figure}{0}

\section{Solution of the eigenvalue equation and estimates of
transverse coherence in the limit of wide electron beam}

\subsection{Basic equation}

   Let us have at the undulator entrance a continuous electron beam
with the current $I_0$, with the Gaussian distribution in energy

\begin{equation}
F({\mathcal E}-{\mathcal E}_{0}) =
\left( 2\pi \langle  (\Delta {\mathcal E})^{2}\rangle   \right)^{-1/2}
\exp \left( - \frac{ ({\mathcal E} - {\mathcal E}_0)^2}
{2\langle  (\Delta {\mathcal E})^{2}\rangle  } \right) \ ,
\label{eq:1.29a}
\end{equation}

\noindent and in a transverse phase plane

\begin{equation}
f(x, x') = (2 \pi \sigma^2 k_{\beta})^{-1}
\exp \left[ - \frac{x^2 + (x')^2/k_{\beta}^2}{2 \sigma^2} \right] \ ,
\label{eq:trans-dist}
\end{equation}

\noindent the same in $y$ phase plane. Here
$k_{\beta} = 1/\beta$ is the wavenumber of betatron oscillations
and $\sigma = \sqrt{\epsilon \beta}$.

The results of this paper can be used in the case of superposition of the
natural undulator focusing and an alternating-gradient external
focusing if the following
condition is satisfied \cite{eigen-general}:

\begin{displaymath}
\frac{L_{f}}{2 \pi \beta} \ll \min \left(1,
\frac{\lambda}{2 \pi \epsilon} \right) \ ,
\end{displaymath}

\noindent where $L_{f}$ is a period of the external focusing structure,
$\beta$ is an average beta-function,
$\epsilon$ is rms emittance of an electron beam, and
$\lambda$ is a radiation wavelength. This condition is met in many
practical situations.

   Using cylindrical coordinates, in the high-gain limit we seek the
solution for a slowly varying complex amplitude of the electric field
of the electromagnetic wave in the form \cite{book}:

\begin{equation}
\tilde{E}(z, r, \varphi ) = \Phi _{nm}(r) \exp (\Lambda z) \left(
\begin{array}{c}
\sin (n\varphi ) \\
\cos (n\varphi )
\end{array}
\right) \ ,
\label{eq:9.1}
\end{equation}

\noindent where $n$ is an integer, $n \ge  0$. For each $n$ there are
many radial eigenmodes that differ by eigenvalue $\Lambda$ and
eigenfunction $\Phi_{nm} (r)$. The integro-differential equation for
radiation field eigenmodes \cite{ming,kim,yu88} taking into account
the space charge effect \cite{eigen-general} can be written in the following
normalized form:

\begin{eqnarray}
& \mbox{} & \left[
\frac{\D^2}{\D \hat{r}^{2}} +
\frac{1}{\hat{r}} \frac{\D }{\D \hat{r}} -
\frac{n^2}{\hat{r}^{2}} +
2\I B \hat{\Lambda} \right]
\Phi_{nm}(\hat{r}) =
-4 \int \limits^{\infty}_{0} \D \hat{r}' \hat{r}'
\left\{
\Phi_{nm}(\hat{r}') \right. \nonumber \\
& \mbox{} & \left. + \frac{\hat{\Lambda }^{2}_{\mathrm{p}}}{2}
\left[
\frac{\D ^2}{\D \hat{r}'^{2}} +
\frac{1}{\hat{r}'} \frac{\D }{\D \hat{r}'} -
\frac{n^2}{\hat{r}'^{2}} +
2\I B \hat{\Lambda} \right]
\Phi_n(\hat{r}') \right\} \nonumber \\
& \mbox{} & \times
\int \limits^{\infty }_{0} \D \zeta \frac{\zeta}{\sin^2
(\hat{k}_{\beta} \zeta)} \exp \left[ -\frac{
 \hat{\Lambda}^2_{\mathrm{T}} \zeta^2}{2} -
(\hat{\Lambda} + \I \hat{C}) \zeta \right] \nonumber \\
& \mbox{} & \times
\exp \left[ -\frac{(1- \I B \hat{k}_{\beta}^2 \zeta/2)
(\hat{r}^2 + \hat{r}'^2)}{\sin^2 (\hat{k}_{\beta} \zeta)}
\right] \nonumber \\
& \mbox{} & \times I_n \left[ \frac{2 (1- \I B \hat{k}_{\beta}^2
\zeta/2) \hat{r} \hat{r}' \cos (\hat{k}_{\beta} \zeta)} {\sin^2
(\hat{k}_{\beta} \zeta)} \right] \ ,
\label{eq:bet-gen-fin}
\end{eqnarray}

\noindent where $I_n$ is the modified Bessel function of the first
kind. The following notations are used here:  $\hat{r} = r/(\sigma
\sqrt{2})$, $B = 2 \sigma^{2}\Gamma \omega /c$ is the diffraction
parameter, $\hat{k}_{\beta} = k_{\beta}/\Gamma$ is the
betatron motion parameter, $\hat{\Lambda }^{2}_{\mathrm{p}} =
2c^{2}(A_{\mathrm{JJ}} \theta _{\mathrm{s}} \sigma \omega )^{-2}$ is
the space charge parameter, $\hat{\Lambda }^{2}_{\mathrm{T}} = \langle
(\Delta {\mathcal E})^2 \rangle/(\rho^2 {\mathcal E}^2)$ is the energy
spread parameter, $\hat{C} = \left[ k_{\mathrm{w}} - \omega/(2c
\gamma_{z}^2) \right]/\Gamma$ is the detuning parameter, $\Gamma =
\left[ A_{\mathrm{JJ}}^2 I_0\omega ^{2}\theta
^{2}_{\mathrm{s}}\left(I_{\mathrm{A}}c^{2}\gamma ^{2}_{z}\gamma
\right)^{-1} \right]^{1/2}$ is the gain parameter, $\rho = c
\gamma_{z}^2 \Gamma/\omega$ is the efficiency parameter, $\omega$ is
the frequency of the electromagnetic wave,
$\theta_{\mathrm{s}}=K_{\mathrm{rms}}/\gamma$, $K_{\mathrm{rms}}$ is
the rms undulator parameter, $\gamma$ is relativistic factor, $\gamma
^{-2}_{z} = \gamma ^{-2}+ \theta ^{2}_{\mathrm{s}}$, $k_{\mathrm{w}}$
is the undulator wavenumber, $I_{\mathrm{A}} =$ 17 kA is the Alfven
current, $A_{\mathrm{JJ}} = 1$ for helical undulator and
$A_{\mathrm{JJ}} = J_0(K_{\mathrm{rms}}^2/2(1+K_{\mathrm{rms}}^2))
- J_1(K_{\mathrm{rms}}^2/2(1+K_{\mathrm{rms}}^2))$ for planar
undulator.  Here $J_0$ and $J_1$ are the Bessel functions of the first
kind.  The space charge effect is included into (\ref{eq:bet-gen-fin})
under the condition $\sigma^2 \gg c^2 \gamma^{2}_{z} /\omega^2$.

\subsection{Exact solution}

   As suggested in \cite{ming} we apply to (\ref{eq:bet-gen-fin}) the
Hankel transformation defined by the following transform pair:

\begin{displaymath}
\bar{\Phi}_{nm} (p) =
\int \limits^{\infty}_{0} \D \hat{r} \hat{r} J_n (p \hat{r})
\Phi_{nm} (\hat{r}) \ , \qquad
\Phi_{nm} (\hat{r}) =
\int \limits^{\infty}_{0} \D p p J_n (p \hat{r})
\bar{\Phi}_{nm} (p) \ .
\end{displaymath}

\noindent Then we
obtain the integral equation for the Hankel transform $\bar{\Phi}_{nm}
(p)$ \cite{eigen-general}:

\begin{eqnarray}
\bar{\Phi}_{nm}(p)
& = &
-\frac{ 1 }{ 2 \I B \hat{\Lambda} - p^2}
\int \limits^{\infty}_{0} \D p' p'
\bar{\Phi}_{nm} (p')
\left[
1 + \frac{\hat{\Lambda }^{2}_{\mathrm{p}}
(2 \I B \hat{\Lambda} - p'^2)}{2} \right] \nonumber \\
& \times &
\int \limits^{\infty }_{0} \D \zeta
\frac{\zeta}
{(1- \I B \hat{k}_{\beta}^2 \zeta/2)^2}
\exp \left[ -\frac{
\hat{\Lambda}^2_{\mathrm{T}} \zeta^2}{2} -
(\hat{\Lambda} + \I \hat{C}) \zeta \right] \nonumber \\
& \times &
\exp \left[ -\frac{p^2 + p'^2}{4(1- \I B \hat{k}_{\beta}^2 \zeta/2)}
\right]
I_n \left[ \frac{ p p' \cos (\hat{k}_{\beta} \zeta)}
{ 2 (1- \I B \hat{k}_{\beta}^2 \zeta/2)} \right] \ .
\label{eq:hank-trans}
\end{eqnarray}

\noindent When the space charge field is negligible,
$\hat{\Lambda }^{2}_{\mathrm{p}} \to 0$, this equation is reduced to
that derived in \cite{ming}.

To solve (\ref{eq:hank-trans}) we discretize it:

\begin{displaymath}
p_i = \Delta (i-\frac{1}{2})  \ , \qquad i =1,2, ...,K \ ,
\end{displaymath}
\begin{displaymath}
p'_j = \Delta (j-\frac{1}{2})  \ , \qquad j =1,2, ...,K \ ,
\end{displaymath}

\noindent where $\Delta$ and $K$ should be chosen such that
the required accuracy is provided. Then we obtain a matrix equation

\begin{displaymath}
\bar{\Phi}_{nm}(i) = M_n (i,j) \bar{\Phi}_{nm}(j) \ ,
\end{displaymath}

\noindent or, $[ M_n - I ] \bar{\Phi}_{nm} = 0$, where $I$ is a unit
matrix. Matrix $M_n$ depends on an eigenvalue $\hat{\Lambda}$ as
well as on the problem parameters:  $B$, $\hat{k}_{\beta}$,
$\hat{\Lambda}^2_{\mathrm{T}}$, $\hat{\Lambda }^{2}_{\mathrm{p}}$, and
$\hat{C}$. The eigenvalues of all radial modes for a given azimuthal
index $n$ can be found by solving the equation $| M_n - I | = 0$.  Then
the calculation of the eigenmodes is straightforward.

This algorithm allows one to find with any desirable
accuracy the eigenvalues and eigenfunctions of a high-gain FEL
including all the important effects: diffraction, betatron motion,
energy spread, space charge, and frequency detuning.
Therefore, it can be considered as a universal tool for
calculation and optimization of high-gain FELs of wavelength range from
infrared to X-ray.

\subsection{Parallel cold beam, large diffraction parameter}

A parallel beam limit is the case when there is no betatron motion, i.e.
$\hat{k}_{\beta} \to 0$. Let us also assume here for the sake of compactness
that the effects of energy spread and space charge can be neglected
($\hat{\Lambda}^2_{\mathrm{T}} \to 0$, $\hat{\Lambda }^{2}_{\mathrm{p}} \to
0$). Equation (\ref{eq:bet-gen-fin}) can then be reduced to

\begin{equation}
\left[
\frac{\D^2}{\D \hat{r}^{2}} +
\frac{1}{\hat{r}} \frac{\D }{\D \hat{r}} -
\frac{n^2}{\hat{r}^{2}} +
2\I B \hat{\Lambda} \right]
\Phi_{nm}(\hat{r}) =
- \frac{2\exp(-\hat{r}^2)}{(\hat{\Lambda} + \I \hat{C})^2} \Phi_{nm}(\hat{r})
\ .
\label{cold-beam-diff}
\end{equation}

To find explicit solutions of the eigenvalue equation in
the limit of large diffraction parameter (more specifically, we require
$B^{1/3} \gg 1$), we use the variational method \cite{ming,ming-deacon}.
We construct a variational
functional from (\ref{cold-beam-diff}):

\begin{eqnarray}
\int \limits_{0}^{\infty} \D \hat{r} \hat{r} \Phi_{nm}(\hat{r})
\left[
\frac{\D^2}{\D \hat{r}^{2}} +
\frac{1}{\hat{r}} \frac{\D }{\D \hat{r}} -
\frac{n^2}{\hat{r}^{2}} +
2\I B \hat{\Lambda} \right]
\Phi_{nm}(\hat{r})
\nonumber \\
=
-\int \limits_{0}^{\infty} \D \hat{r} \hat{r}
\frac{2\exp(-\hat{r}^2)}{(\hat{\Lambda} + \I \hat{C})^2} \Phi_{nm}^2(\hat{r})
\ ,
\label{cold-beam-var}
\end{eqnarray}

\noindent and seek for a solution in the form

\begin{equation}
\Phi_{nm}(\hat{r}) = \hat{r}^n \exp (-a \hat{r}^2) L_m^n (2 a \hat{r}^2) \ ,
\label{trial-laguerre}
\end{equation}

\noindent where $L_m^n$ are associated Laguerre polynomials

\begin{equation}
L_m^n (x)= \frac{1}{m!} \sum_{k=0}^m \frac{m!}{k!} \left( \frac{n+m}{m-k} \right) (-x)^k
\ ,
\label{laguerre}
\end{equation}

\noindent and

\begin{displaymath}
\left( \frac{n}{k} \right) = \frac{n!}{k!(n-k)!}
\end{displaymath}

\noindent is a binomial coefficient.

Equation~(\ref{cold-beam-var}) and the variational condition, $\delta
\hat{\Lambda} / \delta a = 0$, lead to the following two equations for two
unknown quantities, $\hat{\Lambda}$ and $a$:

\begin{equation}
1+n+2m-\frac{\I B \hat{\Lambda}}{a} - \frac{1}{a(\hat{\Lambda} +
\I \hat{C})^2}\left(1-\frac{1+n+2m}{2a}\right) = 0 \ ,
\label{par-1}
\end{equation}

\begin{equation}
\I B \hat{\Lambda} + \frac{1}{(\hat{\Lambda}+\I \hat{C})^2}
\left(1-\frac{1+n+2m}{a}\right) = 0 \ .
\label{par-2}
\end{equation}

\noindent Solving Eqs.~(\ref{par-1}), (\ref{par-2}) in zero order, we get 1-D asymptote
for the eigenvalue equation \cite{book}. The eigenvalue for a growing solution
reaches maximum at $\hat{C}=0$:

\begin{displaymath}
\hat{\Lambda}_0 \simeq \frac{\sqrt{3}+ \I}{2B^{1/3}}
\end{displaymath}

\noindent Then we can find first order correction (in $B^{-1/3}$) to the growth rate
$\re \hat{\Lambda}$, and a mode parameter $a$:

\begin{equation}
\re \hat{\Lambda}_{nm} \simeq \frac{\sqrt{3}}{2B^{1/3}}
\left(1-\frac{\sqrt{2}(1+n+2m)}{3\sqrt{3}B^{1/3}} \right) \ ,
\label{growth-cold}
\end{equation}

\begin{equation}
a \simeq \frac{(\sqrt{3}- \I)B^{1/3}}{2\sqrt{2}} \ .
\label{a-cold}
\end{equation}

Equations~(\ref{trial-laguerre}), (\ref{growth-cold}), and (\ref{a-cold}) are
the solutions for field distributions and growth rates of eigenmodes of a
high-gain FEL with a cold parallel beam in the limit $B^{1/3} \gg 1$. Note that
in \cite{book,opt-comm} the exact solution of the eigenvalue problem for a
parabolic beam density distribution was obtained. The eigenfunctions were
expressed in terms of the confluent hypergeometric function. In this case, in
the limit of a large diffraction parameter the growth rates of eigenmodes are
reduced to (\ref{growth-cold}). The confluent hypergeometric function takes the
form of the associated Laguerre polynomials, so that field distribution is
reduced to (\ref{trial-laguerre}) with the parameter $a$ given in
(\ref{a-cold}). This is not a surprise because in this limit the width of the field distribution
is much smaller than the width of the electron beam ($\re a \simeq B^{1/3} \gg 1$), and the electron
density function behavior is important only near the axis. This behavior is the same (quadratic)
for parabolic and Gaussian distributions. We can also conclude that this asymptotical solution is
valid for any axisymmetric density distribution with the maximum density on axis. We also see that
the variational solution is a good asymptotical method. The attempts to generalize it to the
entire (or, at least wider) parameter range \cite{ming,ming-deacon,ming-high-order}
lead to the loss of accuracy control, although
can give a practically useful fit of the exact solution within some range.

\subsection{Large emittance}

Let us still assume that the space charge effect can be neglected
($\hat{\Lambda }^{2}_{\mathrm{p}} \to 0$). Applying now the variational method
to the Eq.~(\ref{eq:bet-gen-fin}) with the trial functions
(\ref{trial-laguerre}), we obtain for large diffraction parameter:

\begin{eqnarray}
& \mbox{}
1+n+2m-\frac{\I B \hat{\Lambda}}{a} - \frac{1}{a}
\int \limits^{\infty }_{0} \D \zeta
\frac{\zeta}
{1- \I B \hat{k}_{\beta}^2 \zeta/2}
\exp \left[ -\frac{
\hat{\Lambda}^2_{\mathrm{T}} \zeta^2}{2} -
(\hat{\Lambda} + \I \hat{C}) \zeta \right] \nonumber \\
& \mbox{}
\times \left[
1-\frac{(1+n+2m)}{2}
\left(
\frac{a \hat{k}_{\beta}^2 \zeta^2}{1- \I B \hat{k}_{\beta}^2 \zeta/2}+
\frac{1- \I B \hat{k}_{\beta}^2 \zeta/2}{a}
\right)
\right]
= 0 \ ,
\label{all-1}
\end{eqnarray}

\begin{eqnarray}
& \mbox{}
\I B \hat{\Lambda} +
\int \limits^{\infty }_{0} \D \zeta
\frac{\zeta}
{1- \I B \hat{k}_{\beta}^2 \zeta/2}
\exp \left[ -\frac{
\hat{\Lambda}^2_{\mathrm{T}} \zeta^2}{2} -
(\hat{\Lambda} + \I \hat{C}) \zeta \right] \nonumber \\
& \mbox{}
\times \left[
1-\frac{(1+n+2m)(1- \I B \hat{k}_{\beta}^2 \zeta/2)}{a}
\right]
= 0 \ .
\label{all-2}
\end{eqnarray}

The system of equations (\ref{all-1}) and (\ref{all-2}) can be solved
numerically. In the following we neglect the effect of the energy spread,
$\hat{\Lambda}^2_{\mathrm{T}} \to 0$. We also assume that beta-function is
optimized for the highest FEL gain as it happens in practice. Since diffraction
parameter depends on beta-function, it is more convenient go over to other
normalized parameters. Indeed, diffraction parameter can be rewritten as $B = 2
\hat{\epsilon}/\hat{k}_{\beta}$, where $\hat{\epsilon}=2\pi\epsilon/\lambda$.
Then we can go from parameters $(B,\hat{k}_{\beta})$ to
$(\hat{\epsilon},\hat{k}_{\beta})$. After optimizing parameters
$\hat{k}_{\beta}$ and $\hat{C}$, we will find growth rates and eigenfunctions
of all eigenmodes depending on the only parameter, $\hat{\epsilon}$.
Equations~(\ref{all-1}) and (\ref{all-2}) can now be written as
($\hat{\epsilon} \gg 1$):

\begin{eqnarray}
& \mbox{}
1+n+2m-\frac{2\I \hat{\epsilon} \hat{\Lambda}}{a\hat{k}_{\beta} } - \frac{1}{a}
\int \limits^{\infty }_{0} \D \zeta
\frac{\zeta}
{1- \I \hat{\epsilon} \hat{k}_{\beta} \zeta}
\exp \left[- (\hat{\Lambda} + \I \hat{C}) \zeta) \right] \nonumber \\
& \mbox{}
\times \left[
1-\frac{(1+n+2m)}{2}
\left(
\frac{a \hat{k}_{\beta}^2 \zeta^2}{1- \I \hat{\epsilon} \hat{k}_{\beta} \zeta}+
\frac{1- \I \hat{\epsilon} \hat{k}_{\beta} \zeta}{a}
\right)
\right]
= 0 \ ,
\label{emit-kb-1}
\end{eqnarray}

\begin{eqnarray}
\frac{2 \I \hat{\epsilon} \hat{\Lambda}}{\hat{k}_{\beta}} +
\int \limits^{\infty }_{0} \D \zeta
\frac{\zeta}
{1- \I \hat{\epsilon} \hat{k}_{\beta} \zeta}
\exp \left[ - (\hat{\Lambda} + \I \hat{C}) \zeta \right]
\left(
1-\frac{(1+n+2m)(1- \I \hat{\epsilon} \hat{k}_{\beta} \zeta)}{a}
\right)
\nonumber \\
= 0 \ .
\label{emit-kb-2}
\end{eqnarray}

In zero order we find \cite{ming}:

\begin{equation}
\hat{\Lambda}_0 =
\frac{\I \hat{k}_{\beta}}{2 \hat{\epsilon}}
\int \limits^{\infty }_{0} \D \zeta
\frac{\zeta}
{1- \I \hat{\epsilon} \hat{k}_{\beta} \zeta}
\exp \left[ - (\hat{\Lambda}_0 + \I \hat{C}) \zeta \right] \ .
\label{emit-1d}
\end{equation}

\noindent Solving this equation numerically, we find that maximal growth rate

\begin{displaymath}
\re \hat{\Lambda}_0 \simeq \frac{0.37}{\hat{\epsilon}}
\end{displaymath}

\noindent is achieved at the optimal values of $\hat{k}_{\beta} \simeq
0.5/\hat{\epsilon}^2$ and $\hat{C} \simeq 0.4/\hat{\epsilon}$. Note that for
optimal beta-function the diffraction parameter can be expressed as $B \simeq 4
\hat{\epsilon}^3$.

Solving Eqs.~(\ref{emit-kb-1}) and (\ref{emit-kb-2}) in the first order in
$\hat{\epsilon}^{-1}$, we obtain:

\begin{equation}
\re \hat{\Lambda}_{nm} \simeq \frac{0.37}{\hat{\epsilon}}
\left(1-\frac{0.83(1+n+2m)}{\hat{\epsilon}} \right) \ ,
\label{growth-emit}
\end{equation}

\begin{equation}
a \simeq (0.44 - 0.51 \I) \hat{\epsilon} \ .
\label{a-emit}
\end{equation}

\noindent Equations~(\ref{trial-laguerre}), (\ref{growth-emit}), and
(\ref{a-emit}) are the solutions for the field distributions and growth rates of
eigenmodes of a high-gain FEL with optimized beta-function in the limit
of $\hat{\epsilon} \gg 1$.

\subsection{Estimates of transverse coherence}

We can now make a simple estimate of the number of transverse modes, $M$, in
the high-gain linear regime of a SASE FEL operation. The degree of transverse
coherence would then be the inverse number of modes (see (\ref{eq:degcoh-m})):

\begin{equation}
\zeta = \frac{1}{M} = \sigma_{\mathrm{P}}^2 \ ,
\label{deg-mode}
\end{equation}

\noindent where $\sigma_{\mathrm{P}}^2 = \langle (P -\langle P \rangle )^2 \rangle /
\langle P \rangle^2 $ is the relative dispersion of the radiation power.
The field of the electromagnetic wave can be represented as a set of modes, see
(\ref{eq:modes}). In the limit, considered here ($B^{1/3} \gg 1$ or
$\hat{\epsilon} \gg 1$), the modes are orthogonal, and the total power can be
written as\footnote{Ensemble average is meant here.}

\begin{equation}
P_{tot} (\hat{z}) \simeq \sum_{n,m} P_{nm} (\hat{z}) =
2 \sum_{n=0}^{\infty} \sum_{m=0}^{\infty} A_{nm}
\exp (2 \re \hat{\Lambda}_{nm} \hat{z}) - A_{00} \exp (2 \re \hat{\Lambda}_{00} \hat{z})
\ .
\label{wtot}
\end{equation}

\noindent Summation over azimuthal index $n$ is done twice here since for $n \ne 0$ there are two
orthogonal modes that degenerate \cite{book}.
One can also show that (here we assume it without a proof) the factors $A_{nm}$
are the same for all modes in the considered asymptote\footnote{More strictly,
orthogonality of the radial modes with the same azimuthal index, as well as
equality of the factors $A_{nm}$, hold with an accuracy $\hat{\epsilon}^{-1}
\ll 1$. Taking these corrections into account would result in the correction of
the order of $(\re \hat{\Lambda}_{0} \hat{z})^{-1}$ to the number of modes.}.
Since the power of each mode fluctuates in accordance with the negative
exponential distribution (\ref{neg-exp-1}), the dispersion is equal to the
squared average power for each mode. The total dispersion is simply the sum of
dispersions because the modes are independently excited. Thus, the inverse
relative dispersion (or, number of modes) can be calculated as $(\sum
P_{nm})^2/\sum P_{nm}^2$, or explicitly:

\begin{equation}
M \simeq
\frac{\left( 2 \sum \limits_{n=0}^{\infty} \sum \limits_{m=0}^{\infty}
\exp [- 2 N_{g} b (n+2m)] - 1 \right)^2}
{2 \sum \limits_{n=0}^{\infty} \sum \limits_{m=0}^{\infty}
\exp [- 4 N_{g} b (n+2m)] - 1}
 \ ,
\label{mode-ratio}
\end{equation}

\noindent where $N_{g} = \re \hat{\Lambda}_0 \hat{z}$ is a number of field gain lengths along
the undulator, $b=\sqrt{2}/(3\sqrt{3}B^{1/3})$ for a cold parallel beam, and
$b=0.83/\hat{\epsilon}$ for a beam with large emittance and optimized beta-function.
Equation~(\ref{mode-ratio}) is valid when $b \ll 1$ and $N_{g} \gg 1$.

In a particular case when
$1 \ll N_{g} \ll b^{-1}$ the summation in (\ref{mode-ratio}) can be substituted by the
integration. Then for a cold parallel beam we get:

\begin{equation}
M \simeq \frac{27}{2} \left( \frac{B^{1/3}}{N_{g}} \right)^2 \ \ \ \
\mathrm{for} \ \ \ 1 \ll N_{g} \ll B^{1/3} \ .
\label{mode-cold}
\end{equation}

\noindent For a beam with a large emittance and optimized beta-function the
number of modes is

\begin{equation}
M \simeq 1.45 \left( \frac{\hat{\epsilon}}{N_{g}} \right)^2 \ \ \ \
\mathrm{for} \ \ \ 1 \ll N_{g} \ll \hat{\epsilon} \ .
\label{mode-emit}
\end{equation}

We note that applicability region of these estimations is the high-gain linear
regime. Numerical simulations presented in this paper show that the maximum
degree of transverse coherence is achieved already in the nonlinear mode of
operation. Linear analysis, presented here, does not allow to describe this
maximum degree of transverse coherence. However, it can be roughly estimated if
one substitutes $N_{g}$ by the number of field gain lengths at the end of
the linear regime. As an estimate, one can take about 70\% of the number of
field gain lengths required to reach saturation \footnote{Note that saturation
occurs earlier for a larger number of modes. This would give a weak
(logarithmic) correction to the value of the transverse coherence.}. In any
case the asymptotical behavior of the degree of transverse coherence is

\begin{displaymath}
\zeta \propto \frac{1}{\hat{\epsilon}^2}
\end{displaymath}

\noindent in the case of a beam with large emittance and optimized
beta-function.

\end{document}